\documentclass[fleqn,usenatbib]{mnras}

\usepackage{newtxtext,newtxmath}

\usepackage[T1]{fontenc}
\usepackage{ae,aecompl}

\usepackage{graphicx}	
\usepackage{amsmath}	
\usepackage{amssymb}	
\usepackage[flushleft]{threeparttable}
\usepackage{tablefootnote}

\title[Star-formation law in M33]{The star-formation law at GMC scales in M33, the Triangulum Galaxy}

\author[T. G. Williams et al.]{
Thomas G. Williams,$^{1}$\thanks{E-mail: thomas.williams@astro.cf.ac.uk}
Walter K. Gear,$^{1}$
and Matthew W. L. Smith$^{1}$
\\
$^{1}$School of Physics \& Astronomy, Cardiff University, Queens Buildings, The Parade, Cardiff, CF24 3AA, UK
}

\date{Accepted XXX. Received YYY; in original form ZZZ}

\pubyear{2017}

\begin{document}
\label{firstpage}
\pagerange{\pageref{firstpage}--\pageref{lastpage}}
\maketitle

\begin{abstract}
We present a high spatial resolution study, on scales of $\sim$100pc, of the relationship between star-formation rate (SFR) and gas content within Local Group galaxy M33. Combining deep SCUBA-2 observations with archival GALEX, SDSS, WISE, \textit{Spitzer} and submillimetre \textit{Herschel} data, we are able to model the entire SED from UV to sub-mm wavelengths. We calculate the SFR on a pixel-by-pixel basis using the total infrared luminosity, and find a total SFR of $0.17 \pm 0.06\,\rm{M}_\odot$/yr, somewhat lower than our other two measures of SFR -- combined FUV and 24\micron \,SFR ($0.25^{+0.10}_{-0.07}\,\rm{M}_\odot$/yr) and SED-fitting tool MAGPHYS ($0.33^{+0.05}_{-0.06}\,\rm{M}_\odot$/yr). We trace the total gas using a combination of the 21cm H{\sc i} line for atomic hydrogen, and CO(\textit{J}=2-1) data for molecular hydrogen. We have also traced the total gas using dust masses. We study the star-formation law in terms of molecular gas, total gas, and gas from dust. We perform an analysis of the star-formation law on a variety of pixel scales, from 25\arcsec \,to 500\arcsec \, (100pc to 2kpc). At kpc scales, we find that a linear Schmidt-type power law index is suitable for molecular gas, but the index appears to be much higher with total gas, and gas from dust. Whilst we find a strong scale dependence on the Schmidt index, the gas depletion timescale is invariant with pixel scale.

\end{abstract}




\section{Introduction}\label{sec:intro}

An understanding of the processes that govern star-formation within a galaxy is vital to understanding how galaxies form and evolve. Historically, studying these relationships has been limited to the scale of entire galaxies, although this has led to important relations, such as the star-formation law. This law relates the star-formation rate (SFR) of a galaxy to its gas content, describing how efficiently gas is transformed into stars, and thus constraining theoretical models of star-formation. \cite{1959Schmidt} originally suggested a power-law scaling relation between the volume density of SFR to a volume density of gas, i.e.
\begin{equation} \label{schmidt_equation}
\rho_\text{SFR} \propto \rho_\text{gas}^n,
\end{equation}
and this original work by \cite{1959Schmidt} observed a power-law index of $n=2$ for the Milky Way (MW). The first extragalactic measurements of this star-formation law were that of \cite{1969Sanduleak} and \cite{1971Hartwick}, who found a value of $n$ to be $1.84 \pm 0.14$ in the SMC and $3.5 \pm 0.12$ in M31, respectively. However, since most observations of extragalactic objects can only average a surface density along a line-of-sight, more recent studies use a surface, rather than a volume density, i.e. 
\begin{equation} \label{k_s_equation}
\Sigma_\text{SFR} = A\Sigma_\text{gas}^N,
\end{equation}
where $\Sigma_\text{SFR}$ and $\Sigma_\text{gas}$ are the surface densities of SFR and gas, respectively. The current form of this law was studied for a series of $\sim$100 galaxies in the seminal work of \cite{1998Kennicutt}, which found a very tight scaling relation, with $N \sim 1.4$; this ``Kennicutt-Schmidt'' or KS law appears to have a similar $N$ (the so-called Schmidt, or KS index), over a wide range of redshift and environments (see, e.g. \citealt{2012Kennicutt} and references therein). Although this is, necessarily, an oversimplification of a series of complex processes, this indicates that the gas content of a galaxy is a major driver of star-formation, and star-formation is more efficient at higher gas densities. However, these lower resolution studies were unable to determine whether the molecular or total gas content of a galaxy is more strongly correlated with star-formation (e.g. \citealt{2002WongBlitz}). It is also unclear from these works whether star-formation is governed by local processes within star-forming clouds (e.g. \citealt{2005KrumholzMcKee}), or global processes such as cloud-cloud interactions (e.g. \citealt{1986Wyse}). 

One physical interpretation of this empirically derived law is that roughly constant fractions of the total gas present in molecular clouds convert into stars on their free-fall time \citep{1994Elmegreen,2007KrumholzThompson}. This interpretation produces a Schmidt index of 1.5. Another is that the SFR is dictated by the amount of dense molecular gas, with the star-formation law being linear given a constant dense gas fraction. In this case, the traditional superlinear star-formation law is simply an artifact of the variations in this dense gas fraction between the star-forming disk galaxies and starburst galaxies used in studies \citep{2012Lada}. For a series of nearby spiral galaxies, \cite{2008Bigiel} found $N\sim1$ to be a suitable index at sub-kpc scales when considering H$_2$, traced by the $^{12}$CO(\textit{J}=1-0) line -- this would indicate that the molecular gas is simply counting uniform populations of Giant Molecular Clouds (GMCs).

Dense (with number densities, $n>10^4\text{cm}^{-3}$) molecular gas is also a promising tracer of star-formation, as stars are believed to condense out of the dense gas in GMCs \citep{2010Andre,2010Lada}. If this is the case, then we would expect the dense gas mass and SFR to be strongly and linearly correlated. A number of works comparing the far infrared (FIR) luminosity and proxies for this dense gas, such as HCN or HCO$^+$ have found this to be the case (e.g. \citealt{2004GaoSolomona,2004GaoSolomonb,2012GarciaBurillo}). These relationships appear to hold down to $\sim$kpc regions \citep{2015Bigiel,2015Usero}, and large programs are looking to extend these samples \citep{2017Gao}.

With the advent of higher resolution, multi-wavelength surveys of nearby galaxies, our understanding of this star-formation law has improved dramatically. It has been suggested that the KS law would appear to break down on scales similar to that of a GMC ($\sim$10-100pc; e.g. \citealt{2010Onodera,2015Boquien,2017Khoperskov}). \cite{2008Bigiel} claim that the molecular, rather than total gas better correlates with SFR. Work by \cite{2013Ford} suggests that a superlinear $N$ is suitable for the total gas content of M31, whilst a sublinear star-formation law is applicable when considering only molecular gas. \cite{2008Leroy} have found a radial dependence in $N$, with decreasing star-formation efficiency at larger galactocentric radius for a series of 23 nearby galaxies. For apertures targeted on CO and H$\alpha$ peaks in M33, \cite{2010Schruba} find a breakdown at the star-formation law at scales of 300pc, although with increasing aperture size the correlation is restored, which they argue indicates variations between the evolutionary states of GMCs in a galaxy.

M33 is the third massive disk galaxy of our Local Group (behind the MW and M31), and is an excellent laboratory for high-resolution extragalactic studies. M33 is a late-type spiral galaxy located at a distance of 840kpc \citep{1991MadoreFreedman}, and is more face-on than M31 with a moderate inclination of 56\degr \,\citep{1994ReganVogel}. With a large optical extent ($R_{25}$) of 30.8\arcmin \,($\sim$7.4kpc, \citealt{2003Paturel}), M33 is ideally suited for detailed study. Despite being smaller and less massive than the MW, it has a much higher gas fraction, and is actively star-forming throughout its disk \citep{2004Heyer}. It has a high star-formation efficiency \citep{2007Gardan}, with a molecular gas depletion timescale of $1.6-3.2\times 10^8$\,yr, shorter than other local spiral galaxies ($1-3$\,Gyr, \citealt{1998bKennicutt,2002Murgia,2002WongBlitz}). It also has a roughly half-solar metallicity ($12+\log(\text{O/H})=8.36\pm0.04$, \citealt{2008RosolowskySimon}), with a shallow metallicity gradient, making it more analogous to younger or higher redshift galaxies than the MW or M31. Recent work has suggested that following a tidal encounter with M31 \citep{2010McConnachie,2013Wolfe}, the stripped gas now returning to the disk of M33 is fuelling star-formation \citep{2009Putman}. Despite this, the disk of M33 is relatively unperturbed. This is in contrast to the Magellanic Clouds, which are highly disturbed, irregular dwarf galaxies. It is, therefore, particularly noteworthy to study the interplay between the gas and star-formation of M33.

The choice of SFR tracer is critical, as they are sensitive to different timescales and stellar populations.  For instance, \cite{2007Verley} found a SFR of 0.2\,$\rm{M}_\odot$/yr in M33 when considering IR data from \textit{Spitzer}, whilst H$\alpha$ and UV gives a much higher SFR of $0.45\pm0.10\,\rm{M}_\odot$/yr \citep{2009Verley}. Similarly, the adopted tracer of gas and fitting method can have a large impact on the calculated $N$. For M33, \cite{2004Heyer} find a molecular Schmidt index, $N$ of $1.36\pm0.08$, but \cite{2010Verley} find a range of indices ($1.0<N<2.6$), depending on the gas tracer and fitting method employed.

A fundamental assumption of the method used to trace SFR is that the emission arises either directly from young stars, or from their heating of the ISM. UV emission directly tracers the unobscured star-formation from these young stars, and hence is a good tracer of this recent star-formation. However, UV is particularly sensitive to dust attenuation, and so in this work we combine this with 24\micron \,emission to account for re-emission of this dust-absorbed light. 

Another estimate of SFR is the total infrared (TIR) luminosity. This prescription assumes that the dust is heated entirely by young stars. This measure of SFR, however, will miss the starlight that is not absorbed by the dust (e.g. \citealt{2001Hirashita}), and therefore underestimate the SFR. Older stellar populations will also contribute to the heating of this dust, potentially causing an overestimate of the SFR (e.g. \citealt{2008Cortese}).

Finally, we use a third measure of the SFR -- the spectral energy distribution (SED) fitting tool Multi-wavelength Analysis of Galaxy Physical Properties (MAGPHYS, \citealt{2008DaCunha}). This tool fits an SED from a library of models to a series of provided fluxes, and outputs the physical parameters of the fitted model. MAGPHYS allows for a bursty star-formation history, and variations in SFR down to 1Myr. This is particularly important at sub-kpc scales, where the assumption of stationary star-formation may be inappropriate \citep{2009Relano}. It also includes an energy balance between the stellar and dust components of the SED, accounts for filter response, and performs a thorough Bayesian error analysis.

The gas present in the ISM of a galaxy is dominated by hydrogen, both in its atomic (H{\sc i}) and molecular (H$_2$) phases. Whilst the H{\sc i} can be measured directly, via the 21cm line, H$_2$ cannot be traced directly as it is has a low mass, and lacks a dipole moment. Hence, the next most abundant molecule, CO, is commonly used as a proxy, and is traced via its rotational transitions -- in the case of our work, the CO(\textit{J}=2-1) line. 

Alternatively, the gas content of a galaxy can, theoretically, be traced by the cold dust continuum (e.g. \citealt{1983Hildebrand,2012Eales,2012Magdis}). However, the resolution of these dust maps have usually been limited by the resolution of the \textit{Herschel} \citep{2010Pilbratt} Space Observatory. This is particularly true of objects with large angular extents, as ground-based instruments such as SCUBA-2 are poor at recovering large-scale structure due to atmospheric effects. We have developed a technique to combine higher resolution SCUBA-2 data with the larger spatial frequency data of other instruments (Smith et al. in prep), and we present those data for the first time in this work. Using this, we can sample the dust continuum from 100-850\micron \, at an unprecedented spatial scale of 100pc, a factor of $\sim$1.4 better than previous panchromatic galaxy studies \citep{2014Viaene}. This allows us to probe the relationships between star-formation and constituents of the ISM at the scales of individual star-forming regions.

The layout of this paper is as follows: we present an overview of the data used to calculate the SFR, as well as the data processing techniques required to carry out a pixel-by-pixel study of the star-formation law (Section \ref{sec:data}) as well as our methods of calculating SFR at these small scales, and comparisons between them (Section \ref{sec:calc_sfr}). We present an overview of our various methods of tracing the gas within M33 (Section \ref{sec:gas}). We then use this data to study the star-formation law (Section \ref{sec:sf-law}), before our discussion and main conclusions (Section \ref{sec:discussion_conclusions}).

\begin{figure*}
	\includegraphics[width=1.8\columnwidth]{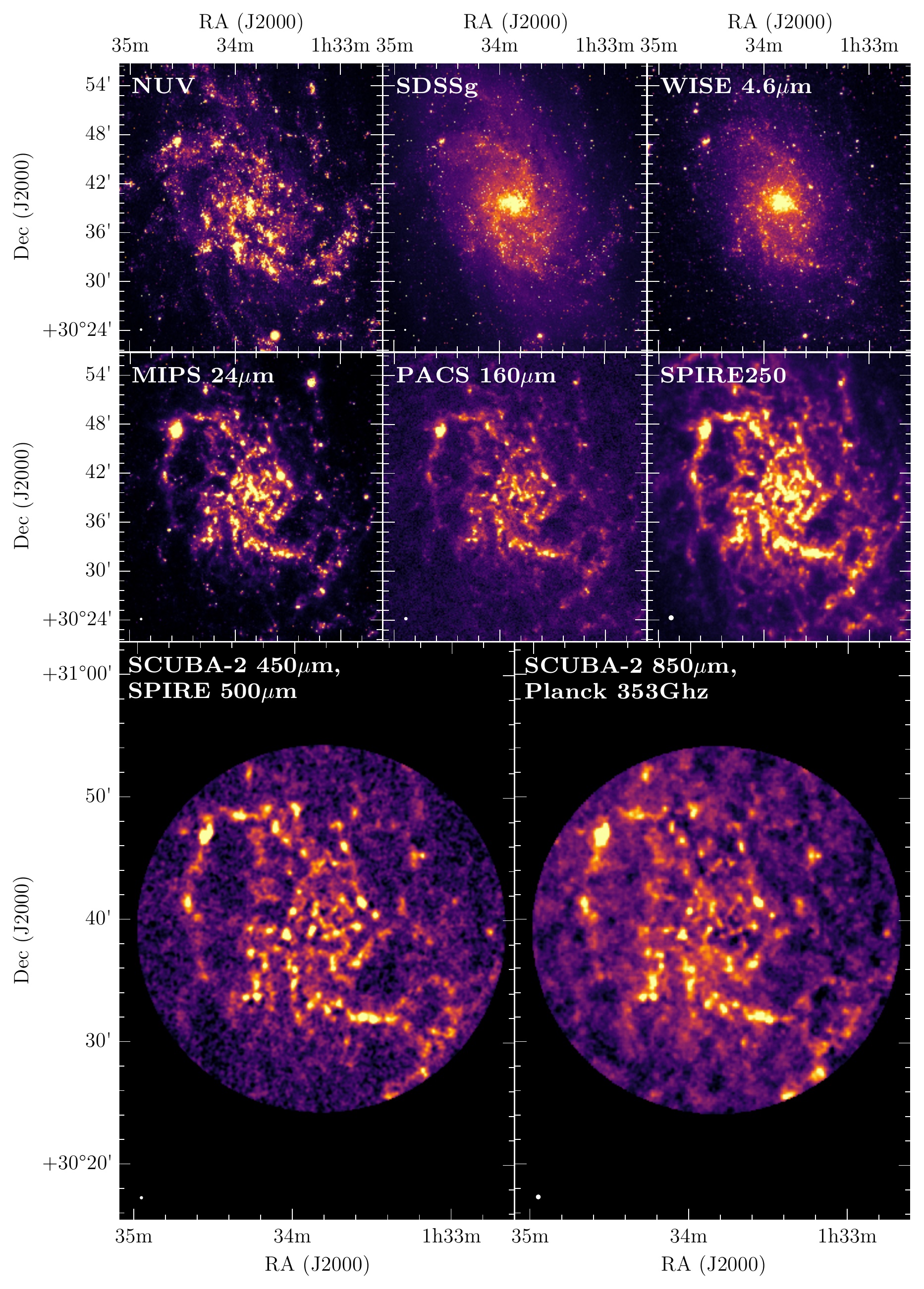}
    \caption{Examples of the data used to calculate SFR in this work. From top left: GALEX NUV \citep{2011Lee} map, SDSS g band mosaic, WISE 4.6\micron \,mosaic, MIPS 24\micron \,\citep{2009Dale} map, PACS 160\micron \,and SPIRE 250\micron \,\citep{2010Kramer} data, and SCUBA-2 data -- 450\micron \,combined with SPIRE 500\micron \,, and 850\micron \,combined with Planck 353GHz data. The SCUBA-2 maps have both been convolved with a 6\arcsec \,Gaussian and cropped to circles of 15\arcmin \,radius, to aid visualisation. The beam for each frame is shown as a solid circle in the bottom left -- in many cases, the beam is negligible compared to the image size.}
    \label{fig:data}
\end{figure*}

\section{Data}\label{sec:data}

In this section, we present an overview of the datasets used in order to probe the star-formation in M33. Particularly for MAGPHYS, it is important to sample the entire galaxy SED from UV to sub-millimetre (sub-mm), so as to provide meaningful constraints on its output parameters. Examples of the images used in this work are shown in Fig \ref{fig:data}.

\subsection{Submillimetre and IR Data}\label{sec:ir_data}

Sub-mm data of M33 at 450 and 850\micron \,was obtained with SCUBA-2 \citep{2013Holland} on the James Clerk Maxwell Telescope (JCMT) between 2017-10-17 and 2017-11-21 under Program ID M17BP003. These 30\arcmin \,PONG maps were taken in band 1 weather (with 225 GHz opacity, $\tau_{225}<0.05$). This data has a resolution of 8\arcsec \,at 450\micron, and 14\arcsec \,at 850\micron. We have also included earlier, public data taken between 2012-07-01 and 2012-07-12 under Program ID M12AC16, taken in marginal band 2/band 3 weather ($0.06<\tau_{225}<0.12$) and acquired from the CADC archive\footnote{\url{http://www.cadc-ccda.hia-iha.nrc-cnrc.gc.ca/en/}}. The final maps have an RMS noise level of 33mJy/beam (450\micron) and 4mJy/beam (850\micron). SCUBA-2 data is unreliable at recovering low spatial frequencies due to the data reduction process due to atmospheric and instrumental variations requiring large-scale filtering in the data reduction process. To mitigate for this we have developed a technique to combine the high spatial frequencies of these data with low spatial frequencies from other telescopes, so as to increase the resolution without losing large-scale structure. For the 450\micron \,data, we set a maximum filter scale of 120\arcsec \,in the data reduction pipeline. The large scale structure is recovered using data from the \textit{Herschel} Spectral and Photometric Imaging Receiver (SPIRE, \citealt{2010Griffin}) instrument, at 500\micron, adjusting the flux accordingly (multiplying the flux by a factor of 1.524, which assumes a fixed dust emissivity index, $\beta$, of 2). The combination of these maps was performed using the \texttt{feather} task in the Common Astronomy Software Applications (CASA) package\footnote{\url{https://casa.nrao.edu/}}. The SCUBA-2 850\micron \,map was treated in much the same way, but instead using 353GHz Planck data, and using a maximum filter scale of 320\arcsec. The maps cover an area larger than 30\arcmin, but the radial dependence on the noise was found to cause artifacts in the feathering process. Hence, for these maps we only use the central 30\arcmin.

Complementing this sub-mm data, we made use of the other two SPIRE bands, at 250 and 350\micron, with a resolution of 18 and 25\arcsec \,respectively. This SPIRE dataset was obtained as part of the \textit{Herschel} M33 extended survey (HerM33es, \citealt{2010Kramer}) open time key project. This project mapped the entirety of M33 with SPIRE, covering a total area of the sky of approximately $70\arcmin\times70\arcmin$. This data was obtained from the Herschel Science Archive\footnote{\url{http://archives.esac.esa.int/hsa/whsa}}, utilising the Standard Product Generation (SPG) software v14.1.0. These maps have been calibrated for extended sources, and we include the small beam correction recommended in the SPIRE handbook \footnote{\url{http://herschel.esac.esa.int/Docs/SPIRE/spire_handbook.pdf}}. The RMS noise levels of this SPIRE data are 14.1, 9.2, and 8mJy/beam at 250, 350 and 500\micron, respectively.

The HerM33es project also mapped this same $70\arcmin\times70\arcmin$ at 100 and 160\micron \,with the Photoconductor Array Camera and Spectrometer (PACS, \citealt{2010Poglitsch}). This data has a resolution of 7.7 and 12\arcsec\,respectively, with an RMS noise level of 2.6mJy pixel$^{-1}$ (100\micron) and 6.9 mJy pixel$^{-1}$ (160\micron). This data was processed using the SPG software v14.2.0, using the \texttt{JScanam} map-maker. Again, these maps are calibrated for extended sources. 

Our first source of near infrared (NIR) data is from the Wide-field Infrared Survey Explorer (WISE\footnote{\url{http://irsa.ipac.caltech.edu/Missions/wise.html}}, \citealt{2010Wright}). These images have wavelengths of 3.4, 4.6, 12 and 22\micron \,with FWHM of 6.1, 6.4, 6.5 and 12\arcsec \,respectively. We used \texttt{Montage}\footnote{\url{http://montage.ipac.caltech.edu}} to mosaic together various frames from the AllWISE data release, incorporating both the WISE cryogenic and NEOWISE \citep{2011Mainzer} post-cryogenic surveys. \texttt{Montage} also matches background levels between each frame, so that overlaps between frames match as closely as possible. However, this is not a background subtraction, so to allow us to adequately model the sky, we created a 3$\deg^2$ mosaic of M33 in each band, to ensure we had a sufficiently large amount of sky to model.

Additional IR data was obtained by the \textit{Spitzer} Infrared Array Camera (IRAC, \citealt{2004Fazio}), as part of the Local Volume Legacy (LVL, \citealt{2009Dale}) Survey\footnote{\url{http://irsa.ipac.caltech.edu/data/SPITZER/LVL/}}. We used data from IRAC taken at 3.6, 4.5, 5.8 and 8\micron. The resolution of these data are $\sim2\arcsec$, and cover approximately $90\arcmin\times60\arcmin$. Along with IRAC data, we also made use of the Multiband Imaging Photometer for \textit{Spitzer} (MIPS, \citealt{2004Rieke}) data, taken again as part of the LVL. This data covers approximately $130\arcmin\times80\arcmin$, and is at 24 and 70\micron, with a resolution of 6 and 18\arcsec \,respectively. The overlap between several \textit{Spitzer} and WISE bands improves sampling of the mid-infrared (MIR) SED, reducing the dependence of the fit on a single point.

\subsection{UV and Optical Data}\label{sec:uv_data}

The UV data used in this work comes from the \textit{Galaxy Evolution Explorer} (GALEX, \citealt{2005Martin}), obtained by \cite{2005Thilker}. Data was obtained for both the FUV (1516\AA) and NUV (2267\AA), covering a circular area of radius $\sim$36\arcmin. The angular resolution of this data is 4.2\arcsec \,and 5.3\arcsec \,for the FUV and NUV, respectively, and with $\sim$3ks exposures, typical $1\sigma$ RMS flux sensitivities are $\sim$28 AB mag arcsec$^{-2}$.

The optical data used in this study comes from the Sloan Digital Sky Survey (SDSS\footnote{\url{https://dr13.sdss.org/home}}, \citealt{2000York}). Using only primary frames from the SDSS DR13 \citep{2015Alam}, a mosaic of 3$\deg^2$ was created, allowing enough sky to accurately model the background. The SDSS data was mosaicked together using \texttt{Montage} for all five bands -- u (3543\AA), g (4770\AA), r (6231\AA), i (7625\AA), and z (9134\AA).

\subsection{Data Preparation}

\begin{table}
	\centering
	\caption{Calibration uncertainty for each pixel.}
	\label{tab:calib_uncert}
	\begin{tabular}{c|c|c}

		\hline
		Telescope & Calibration Uncertainty & Reference\\
		\hline
		\rule{0pt}{2.6ex}
		GALEX FUV & 5\% & 1\\
	    GALEX NUV & 3\% & 1\\
        SDSS u & 2\% & 2\\
		SDSS g,r,i,z & 1\% & 2\\
        IRAC-1 & 10\% & 3\\
        IRAC-2 & 10\% & 3\\
        IRAC-3 & 15\% & 3\\
        IRAC-4 & 15\% & 3\\
        MIPS 24$\mu\text{m}$ & 4\% & 4\\
        MIPS 70$\mu\text{m}$ & 5\% & 5\\
        WISE W1 & 2.4\% & 6\\
        WISE W2 & 2.8\% & 6\\
        WISE W3 & 4.5\% & 6\\
        WISE W4 & 5.7\% & 6\\
        PACS & 5\% & 7\\
        SPIRE & 5.5\% & 8\\
        SCUBA-2 450$\mu\text{m}$ & 12\% & 9\\
        SCUBA-2 850$\mu\text{m}$ & 8\% & 9\\
        Planck 353GHz & 3\% & 10\\
		\hline		
        \multicolumn{3}{p{\linewidth}}{\textbf{References:} 1) \cite{2007Morrissey}; 2) \cite{2008Padmanabhan}; 3) \cite{2009Dale}; 4) \cite{2007Engelbracht}; 5) \cite{2007Gordon}; 6) \cite{2011Jarrett} 7) PACS Handbook\tablefootnote{\url{http://herschel.esac.esa.int/Docs/PACS/pdf/pacs_om.pdf}}; 8) SPIRE Handbook; 9) \cite{2013Dempsey}; 10) \cite{2014Planck} }
	\end{tabular}
\end{table}

Incorporating data from a variety of sources requires careful consideration so that meaningful comparisons can be drawn on a pixel-by-pixel basis. Hence, it was necessary to process the dataset, so as to make it homogeneous, and we give a description of that process here.

\subsubsection{Background Subtraction}\label{sec:background_subtraction}

For each frame, we performed a background subtraction. Depending on the background, we employed a variety of methods to achieve this. Before this background subtraction process, we also converted all of the data into units of Jy/px, if it was required.

\textbf{GALEX:} The average background for the GALEX frame was found to be 0, with no clear gradient, so no background subtraction was applied.

\textbf{SDSS:} Due to the mosaicking process, the SDSS frames had a varying non-zero background. In order to remove this background, M33 was masked (using an ellipse of $80\arcmin \times 60\arcmin$) before fitting and subtracting a 2-dimensional polynomial. This reduced the background variation in the image to the order of a few percent, consistent with, e.g., \cite{2014Corbelli}.

\textbf{\textit{Spitzer}, WISE, \textit{Herschel}:} For the \textit{Spitzer}, WISE and \textit{Herschel} frames, the background was constant throughout the image, so in these frames a median background was subtracted using a $3\sigma$ clipped median after masking all sources with a signal-to-noise (S/N) \textgreater ~2.

\textbf{SCUBA-2:} The SCUBA-2 data reduction process performs an iterative sky modelling and subtraction procedure \citep{2013Chapin}, so no further sky subtraction was performed.

Following background subtraction, we apply a Galactic extinction correction for frames with central wavelengths shorter than 4.6\micron \,(the WISE-2 band). We use the method prescribed by \cite{2011SchlaflyFinkbeiner} for the central position of M33, provided by the IRSA DUST\footnote{\url{http://irsa.ipac.caltech.edu/applications/DUST/}} service. Due to the large angular extent of M33, the extinction correction can vary by a significant amount (14\% in the FUV and NUV frames, for example). We included this variation in our uncertainty treatment (Section \ref{sec:uncertainties}).

\subsubsection{Star Masking}

It was also necessary to mask any flux contamination from foreground stars. We masked stars using a comparison of the UV fluxes, as described by \cite{2008Leroy}. Using \texttt{SExtractor} \citep{1996Bertin}, we found the positions of all $5\sigma$ detections in the NUV. An optimal aperture size was then calculated for each of these detections using a similar method to \cite{2014Viaene}, by calculating a radius for each detection where the flux at that radius dropped below two times the local background level. These apertures were then placed in the FUV maps and the flux within each calculated. \cite{2008Leroy} found that foreground stars have an NUV-to-FUV flux ratio of $\gtrsim 15 \pm 5$. Upon visual inspection, we found that a ratio 15 was insufficient to mask all foreground stars, and so opted instead for a value of 10. These stars were subsquently masked in all frames up to $\sim$20\micron, after which the foreground star emission was no longer an issue. Of the $\sim$6000 sources detected with \texttt{SExtractor}, around 200 were masked.

\subsubsection{Convolution and Regridding}

In order to make comparisons on a pixel-by-pixel basis, it was necessary to match the data to a common resolution and pixel scale. In order to do this, we made use of the convolution kernels\footnote{\url{http://www.astro.princeton.edu/~ganiano/Kernels.html}} provided by \cite{2011Aniano} in order to achieve a common resolution. In this case, we match everything to the PSF of the SPIRE 350\micron \,data, which has a FWHM of 25\arcsec. We regridded all of the data to a common pixel scale of 25\arcsec \,(corresponding to a spatial scale of 100pc), which ensured that we could safely assume each pixel to be statistically independent.  For an ellipse of $60\arcmin\times70\arcmin$, this corresponded to 19004 pixels.

\subsection{Uncertainties}\label{sec:uncertainties}

There were a number of uncertainties to take into account for each pixel. In areas of high S/N, the calibration error ($\sigma_\text{cal}$) of the instrument dominates. We considered these calibration errors, which are summarised in Table \ref{tab:calib_uncert}. In the case of the SCUBA-2 maps combined with other, lower resolution maps, we take the calibration error as the sum of the two relevant uncertainties in quadrature. We also included uncertainties from the varying Galactic extinction correction ($\sigma_\text{ext}$) due to the large angular extent of M33. The GALEX data is most affected by this, with a scatter of 14\%. We also considered the background variation ($\sigma_\text{bg}$) in each frame. For this, we took the standard deviation of the background, having masked any sources greater than $2\sigma$. In the case of the SDSS frames, this error also incorporated any remaining large-scale residuals due to the mosaicking process. Finally, for the GALEX and SDSS frames an uncertainty arose from the small numbers of photons incident at these wavelengths ($\sigma_\text{poiss}$). These errors are Poissonian. To estimate these errors, the flux was converted back into a count number and the square root of this count converted into a flux to give an error. The total uncertainty for each pixel was given by
\begin{equation}
\sigma_\text{total} = \sqrt{\sigma_\text{cal}^2+\sigma_\text{bg}^2+\sigma_\text{ext}^2+\sigma_\text{poiss}^2}
\end{equation}

\section{Calculating SFR}\label{sec:calc_sfr}

\begin{figure*}
	\includegraphics[width=2\columnwidth]{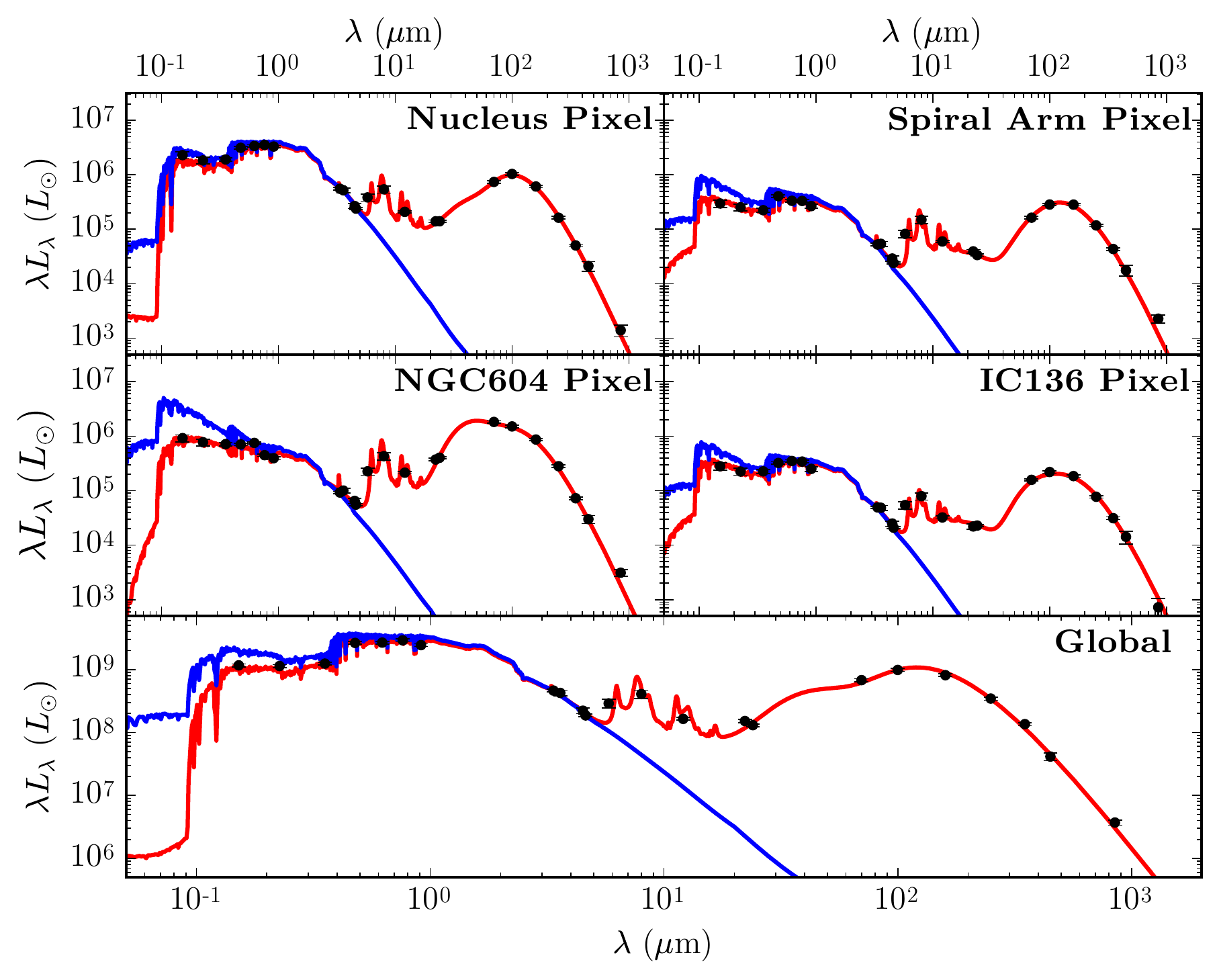}
    \caption{Example MAGPHYS SEDs for single pixels. The blue line represents the unattenuated SED, the red line the best fit to the data (i.e. the dust attenuated SED). From left to right, there is a pixel from within the nucleus (R < 0.5kpc) of M33, a pixel from the northern spiral arm, a pixel from within NGC604 (an H{\sc ii} region) and a pixel from within IC136 (a stellar association). The lowermost panel shows the global SED of M33.}
    \label{fig:example_seds}
\end{figure*}

\begin{figure*}
	\includegraphics[width=2\columnwidth]{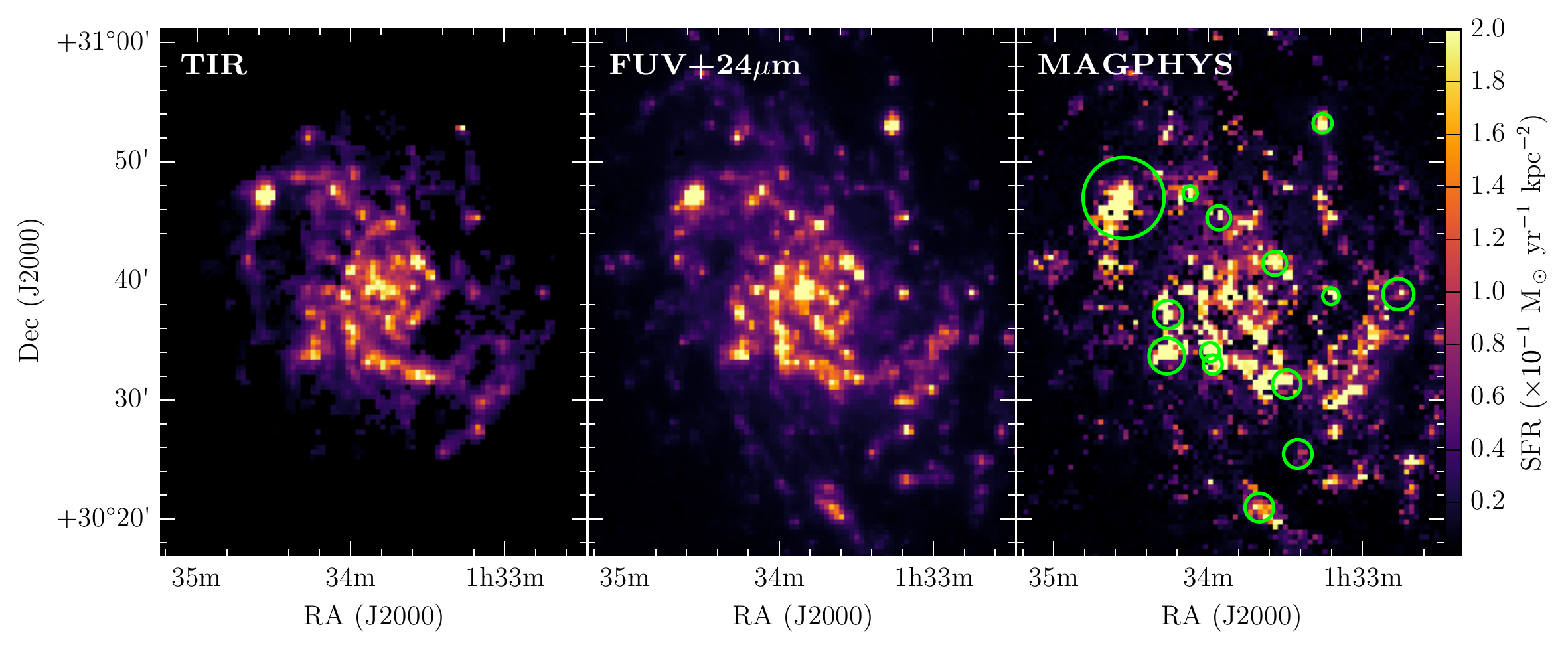}
    \caption{From left to right, SFR density maps from TIR luminosity found by integrating modified blackbody fits from 3-1100\micron, SFR from a combination of FUV+24\micron \,data, and median-likelihood estimates of SFR from MAGPHYS. Particularly bright pixels in the MAGPHYS map are due to recent starbursts, producing an SFR $\sim$10 times higher than the average pixel value, and tend to be associated with H{\sc ii} regions and stellar clusters (a selection of which are shown as green circles). All maps have pixel sizes of 25\arcsec.}
    \label{fig:sfr_maps}
\end{figure*}

Different tracers of SFR are subject to different systematic effects, such as dust attenuation and the impact of older stellar populations. To ensure that our pixel-by-pixel SFR measurements are robust, we have compared three different SFR calibrations.

\subsection{Total Infrared Luminosity}

We first calculated the SFR from the integrated TIR luminosity. TIR luminosity effectively captures the bolometric luminosity of the dust-obscured stellar population, and so traces the starlight absorbed by the dust in a galaxy. Two important assumptions for TIR luminosity tracing the total SFR of a galaxy are that all of the emission from the young stellar population is absorbed by dust, and that the dust heating is exclusively from these young stars. Both of these assumptions are oversimplifications for galaxies (see, e.g. \citealt{2001Hirashita,2008Cortese}), and whilst on the scale of integrated galaxies appear to balance each other out (e.g. \citealt{2002Kewley}), on a pixel-by-pixel basis these assumptions may not hold true. However, the TIR SFR is a useful diagnostic, and informs us both about the stellar population of a galaxy, as well as its dust content. In this work, we have used the calibration given by \cite{2012Kennicutt}, using values from \cite{2011Hao} and \cite{2011Murphy}, integrating the greybody fit from 3-1100\micron:
\begin{equation}
\log_{10}(\text{SFR}_\text{TIR}) = \log_{10}(L_\text{TIR})-43.41,
\end{equation}
where $L_\text{TIR}$ is in ergs s$^{-1}$. This assumes the default IMF of Starburst99 \citep{1999Leitherer}, the broken \cite{2001Kroupa} power law with a maximum mass of $120\,\text{M}_\odot$, and a time-scale of $\sim$100\,Myr. Including an error of 30\% to estimate an uncertainty in the IMF and amount of dust attenuation, we found a TIR SFR of $0.17\pm0.06\,\text{M}_\odot$/yr.

\subsection{FUV+24\micron}

The TIR SFR misses the unattenuated component of the SFR, which can vary dramatically in different environments. Particularly significantly for M33, the unattenuated component can be a major proportion of the total SFR in low metallicity environments (e.g. \citealt{2001Hirashita}). One effective method of overcoming this is to combine IR measurements tracing obscured star-formation with UV emission that measures the unattenuated star-formation. In this work, we have elected to combine the GALEX FUV and MIPS 24\micron \,data. FUV emission traces unobscured star formation over a timescale of $\sim$10-100\,Myr (e.g. \citealt{1998Kennicutt,2005Calzetti}), whilst the 24\micron \,emission traces emission from small dust grains heated by starlight over a timescale of $\sim$10\,Myr (e.g. \citealt{2005Calzetti,2007Calzetti}). We use the SFR density prescription of \cite{2008Leroy}:
\begin{equation}\label{eq:fuv_24_micron}
\Sigma_{\text{SFR}} = 8.1\times10^{-2}I_{\text{FUV}}+3.2^{+1.2}_{-0.7}\times10^{-3}I_{24}
\end{equation}
where $\Sigma_{\text{SFR}}$ is in units of $\text{M}_\odot\,\text{kpc}^{-2}\,\text{yr}^{-1}$, $I_{\text{FUV}}$ and $I_{\text{24}}$ are intensities in units of MJy/sr. Again, we have assumed the default Starburst99 settings, and a time-scale of 100\,Myr. Some of the emission at these wavelengths may be due to an older stellar population (e.g. \citealt{2009Kennicutt}), indicated by a correlation between these bands and the 3.6\micron \,data. To correct for this, we remove this contribution using
\begin{equation}
I_\text{FUV, corr} = I_\text{FUV} - \alpha_\text{FUV} I_{3.6},
\end{equation}
\begin{equation}
I_\text{24, corr} = I_\text{24} - \alpha_\text{24} I_{3.6},
\end{equation}
where $\alpha_\text{FUV} = 3\times10^{-3}$ and $\alpha_\text{24} = 0.1$ \citep{2008Leroy}. This correction has the effect of reducing the total SFR by $0.01 \text{M}_\odot$/yr. Using this method, we find a total SFR of $0.25^{+0.10}_{-0.07}\,\text{M}_\odot$/yr, somewhat higher, but still consistent with the TIR SFR.

\subsection{MAGPHYS}

Finally, we calculate the SFR using MAGPHYS. MAGPHYS fits an SED from a large library of optical and IR models with known, physically motivated input parameters. It finds the best fit to the data in each case, and outputs the physical parameters of these fits, as well as modelling uncertainties upon them. For the optical models, MAGPHYS assumes a \cite{2003Chabrier} initial mass function (IMF), which it evolves using the \cite{2003BruzualCharlot} stellar population synthesis (SPS) model, and has a star-formation history (SFH) resolution of 1\,Myr. Dust obscuration is calculated using the model of \cite{2000CharlotFall}. The total MAGPHYS SFR as calculated from the integrated flux across the galaxy is $0.33^{+0.05}_{-0.06}M_\odot$/yr. These uncertainties only take into account uncertainties on the flux, so are likely an underestimate of the true error. This value for SFR is consistent with work by \cite{2009Verley}, but higher than both the total SFR calculated using FUV+24\micron \, and from TIR luminosity. However, whilst MAGPHYS has a parameter space suitable for integrated galaxies, a single pixel in M33 is far outside this space in terms of flux. We artificially increased each flux by a factor of $10^4$ to put it within MAGPHYS parameter space. Most quantities from MAGPHYS tend to be ratios and so are not affected by this scaling -- the four that scale with flux do so linearly, and are the SFR, dust mass ($\textrm{M}_{\text{dust}}$), stellar mass ($\textrm{M}_{\text{star}}$) and dust luminosity ($L_{\text{dust}}$).

One important feature of MAGPHYS is that it enforces an energy balance, where all attenuated light is re-emitted by the dust. Whilst this may hold true for a whole galaxy, the light from neighbouring regions may have an impact on the pixel in question in these sub-kpc regions. If this is the case, there may be an offset between the values calculated on a per-pixel basis, and those on a global scale. To test this, we also calculated the SFR from the sum of the individual pixels, giving us a value of $0.33\pm 0.10\,\text{M}_\odot$/yr. This indicates that MAGPHYS is suitable for pixel-by-pixel fitting, despite its original intent for galaxy-scale SED fits.

A benefit of using MAGPHYS is that the entire range of data can be used, regardless of the errors on each individual point. The fitted parameters can then be filtered a posteriori, based on how well constrained they are. MAGPHYS gives a probability distribution function (PDF) for each parameter, and the width of this PDF indicates how well constrained each parameter is. MAGPHYS provides in its outputs the 2.5th, 16th, 50th, 84th, and 97.5th percentiles for each parameter, we calculated a relative uncertainty for each pixel -- any pixels which did not satisfy
\begin{equation}
0.5\times\frac{p_{86}-p_{16}}{p_{50}} < 0.32
\end{equation}
where $p_x$ is the $x$th percentile, were removed. This method of pixel filtering has previously been employed by \cite{2014Viaene}, who found it a sufficient cut to remove any broad or bimodal (i.e. poorly constrained) PDFs. We have used this method of filtering pixels in the proceeding analysis, rather than a more traditional S/N cut on the map, leaving us with 6,574 pixels. In this work, rather than using the SFR for the best fit model, we used the median-likelihood estimate given by the PDF. Some example SEDs, as well as the global SED can be seen in Fig. \ref{fig:example_seds}.

\subsection{SFR Comparisons}

\begin{figure}
\begin{center}
\includegraphics[width=\columnwidth]{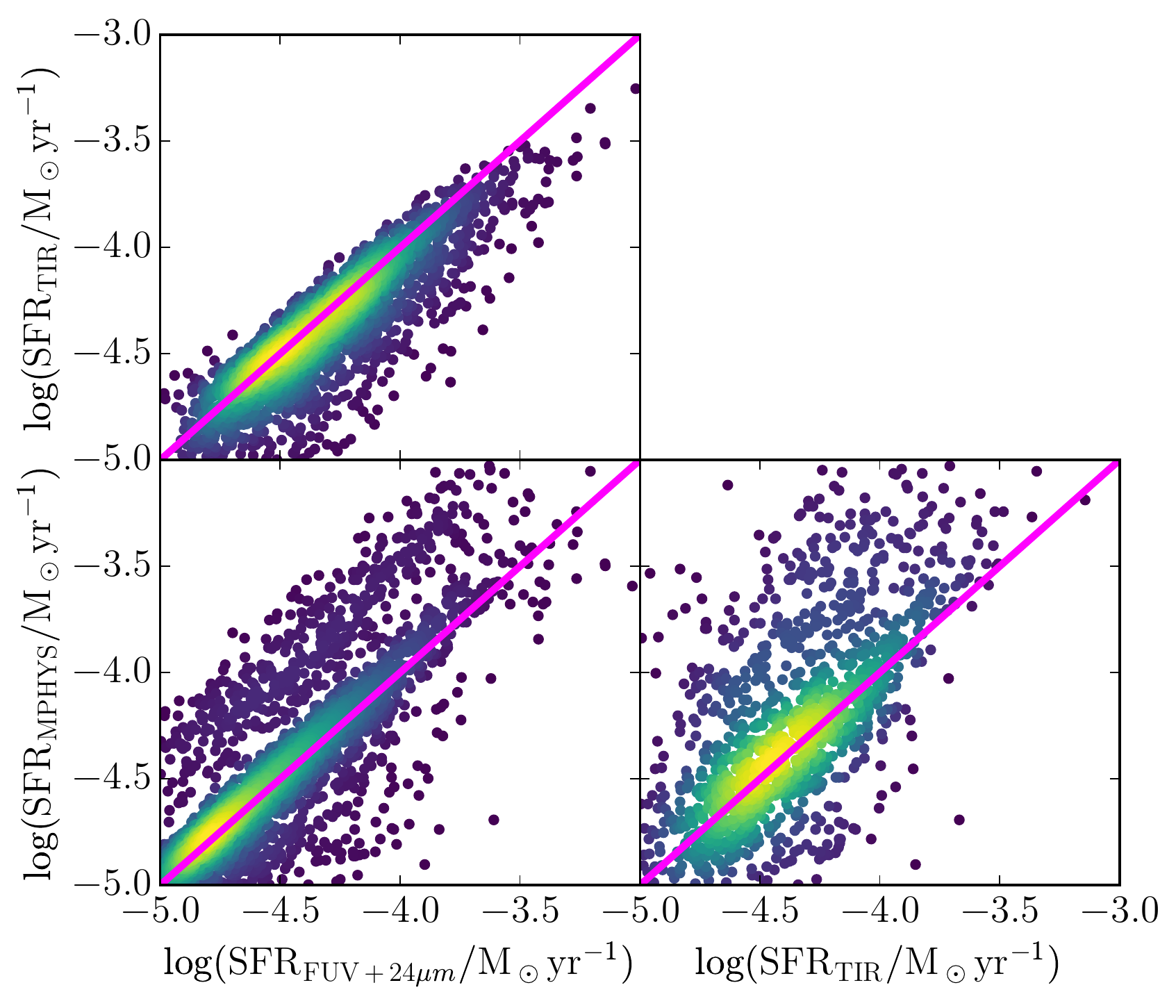}
\caption{Comparisons for single pixel fits between \textit{left,top}: SFR as calculated by the TIR luminosity and by FUV+24\micron \,data. \textit{Left, bottom}: SFR as outputted from MAGPHYS and as calculated by a combination of FUV and 24\micron \,data. \textit{Right, bottom}: SFR parameter from MAGPHYS and calculated from the TIR luminosity. The one-to-one relation is shown as a fuchsia line. Blue points indicate a lower point density, yellow a higher point density.}
\label{fig:sfr_comparisons}
\end{center}
\end{figure}

\begin{table}
 \caption{Comparisons between the three SFR tracers. For the TIR and FUV+24\micron \,maps, only pixels with S/N\textgreater$5\sigma$ have been considered. In the case of MAGPHYS, the filtered map is used. Given are the RMS scatter of the points ($\sigma$) and the median offset from the 1-1 relation [$\Delta$log (SFR)].}
 \label{tab:sfr_comparisons}
 \begin{tabular}{ccccc}
  \hline
  & \multicolumn{2}{c}{SFR$_\text{FUV+24\micron}$} & \multicolumn{2}{c}{SFR$_\text{TIR}$}\\
   & $\sigma$ & $\Delta$log(SFR) & $\sigma$ & $\Delta$log(SFR) \\
   \hline
  SFR$_\text{TIR}$ & 0.16 & -0.05 & - & - \\
  SFR$_\text{MAGPHYS}$ & 0.25 & 0.04 & 0.34 & 0.08 \\
  \hline				
 \end{tabular}
\end{table}

\begin{figure*}
	\includegraphics[width=2\columnwidth]{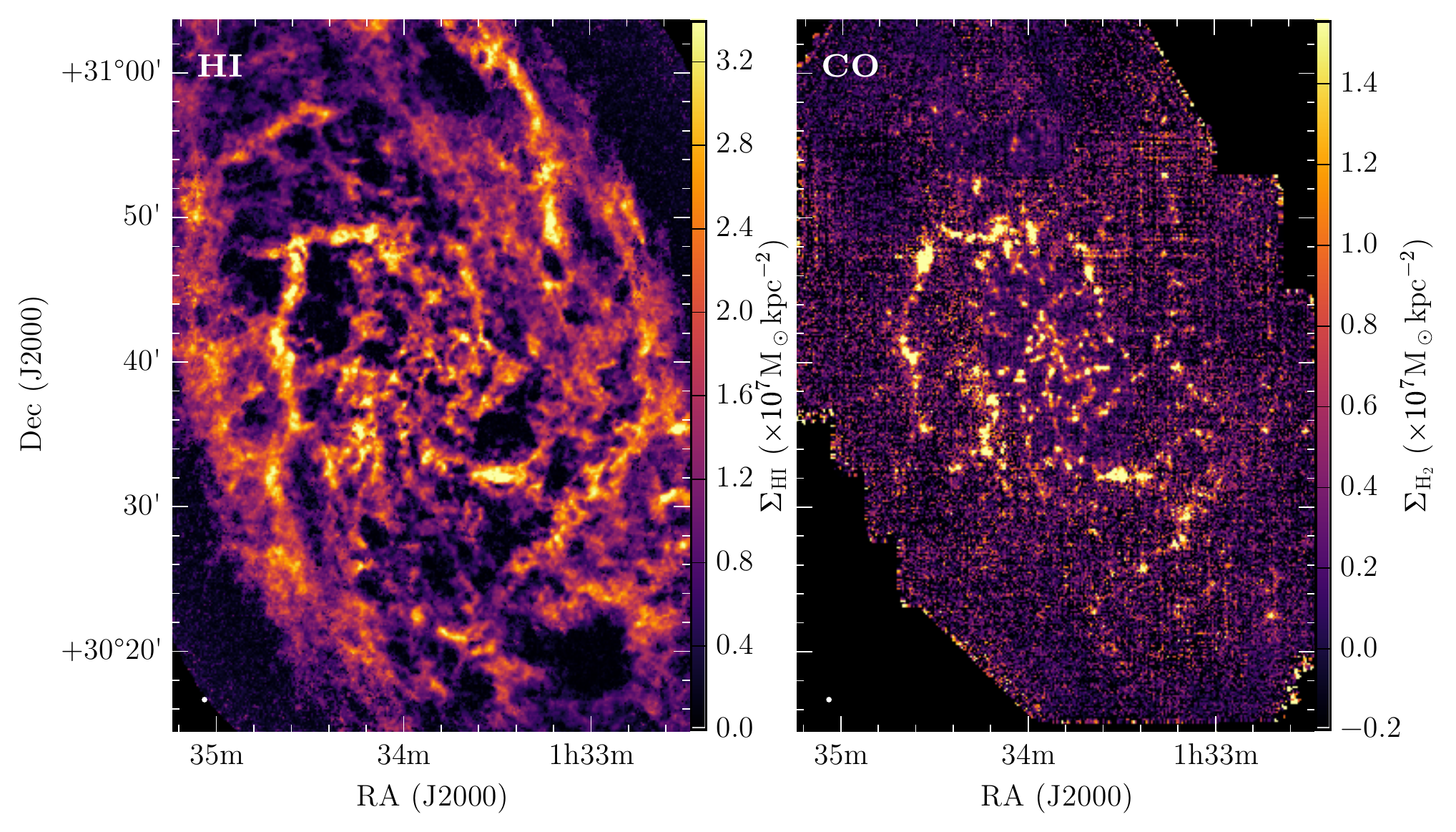}
    \caption{\textit{Left}: Atomic gas mass surface density calculated from integrated H{\sc i} emission \citep{2010Gratier}. \textit{Right}: Molecular gas mass surface density calculated from integrated CO(\textit{J}=2-1) emission \citep{2010Gratier,2014Druard}. The beam is indicated as a white circle in the lower left.}
    \label{fig:gas_data}
\end{figure*}

We find that the lowest SFR calculated is from the TIR luminosity, at $0.17\pm0.06\,\text{M}_\odot$/yr. Including the unattenuated component of the starlight increases the SFR to $0.25^{+0.10}_{-0.07}\,\text{M}_\odot$/yr using FUV+24\micron \,, and even higher from MAGPHYS ($0.33^{+0.05}_{-0.06}\,\text{M}_\odot$/yr). This lack of dust attenuation is highlighted in Fig. \ref{fig:example_seds} -- the stellar component dominates over the dust, as highlighted by the small offset between the unattenuated and attenuated lines, especially within the nucleus. That this unattenuated component accounts for $\sim$50\% of the SFR highlights the importance of the inclusion of the unattenuated component, particularly in low-metallicity or dust-poor galaxies.

As well as a global SFR, we have also calculated the SFR on a per-pixel basis, and these maps can be seen in Fig. \ref{fig:sfr_maps}, with a comparison of these three tracers of SFR in Fig \ref{fig:sfr_comparisons}. In the cases of the TIR and FUV+24\micron \,SFR map, we compare only the pixels with S/N $>5\sigma$. The three broadly agree -- the RMS scatter ($\sigma$) and median offset from the 1-1 relation [$\Delta$log(SFR)] is summarised in Table \ref{tab:sfr_comparisons}. However, there is a population of pixels with higher SFRs given by MAGPHYS than the other two tracers, and this is reflected in a much higher scatter. In the cases where the MAGPHYS SFR is significantly higher, this is due to MAGPHYS injecting a recent starburst. Both the SFR from TIR luminosity and FUV+24\micron \,assume continuous star-formation over the last 100\,Myr. At sub-kpc resolutions, star-formation may vary over timescales of a few Myr \citep{2009Boselli}. We find that these areas of bursty star-formation tend to be associated with H{\sc ii} regions and stellar associations (a sample of which are shown in the rightmost panel of Fig \ref{fig:sfr_maps}). As the spectra of H{\sc ii} regions strongly resemble those of starburst galaxies (e.g. \citealt{1997Ho}), it is not surprising that MAGPHYS has treated them as such. Additionally, UV and FUV spectroscopy has shown that at least 2 populations of stars exist within the nucleus of M33, and that star-formation occurred within the nucleus $\sim$40Myr ago \citep{2002Long}, so this injected starburst within the nucleus of M33 is believable. Overall, these tracers of SFR show very similar characteristics on a pixel-by-pixel level, but as MAGPHYS uses all available data, provides an energy balance, and can take into account recent bursts of star-formation, we elect to use this tracer going forwards.

\section{Gas}\label{sec:gas}

\subsection{Atomic Gas}

\begin{figure*}
	\includegraphics[width=2\columnwidth]{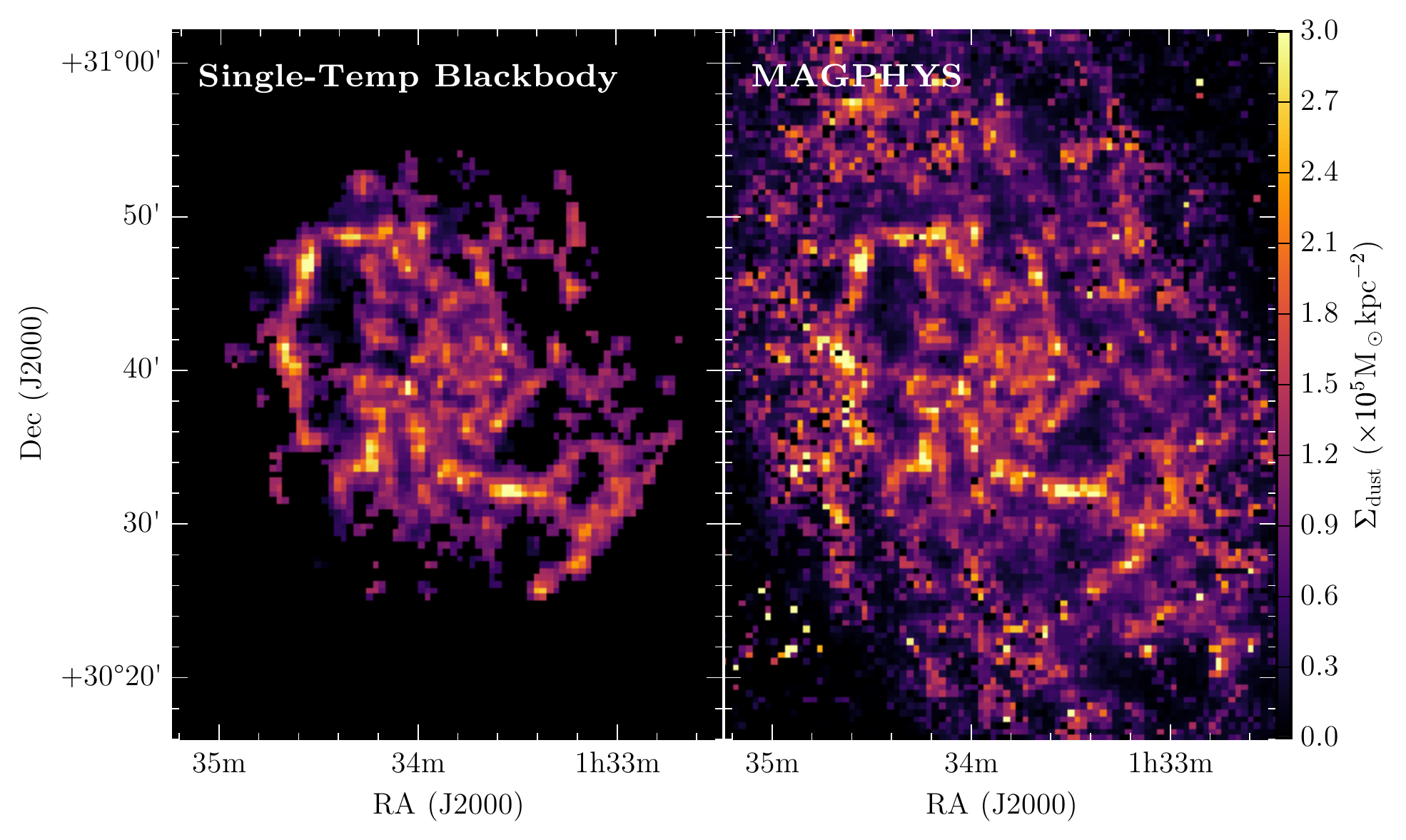}
    \caption{\textit{Left}: Dust mass surface density from one-temperature greybody fit. \textit{Right}: Dust mass surface density from MAGPHYS.}
    \label{fig:dust_data}
\end{figure*}

For studying the atomic hydrogen, archival VLA\footnote{\url{https://science.nrao.edu/facilities/vla/archive/index}} B, C, and D array data for the 21cm line was reduced by \cite{2010Gratier}. The data used was an integrated intensity map in units of K km/s, with a spatial resolution of 12\arcsec \,($\sim$50pc at the distance of M33). The original data cube has a spectral resolution of 1.27 km/s. This data can be seen in the left panel of Fig \ref{fig:gas_data}. From the H{\sc i} 21cm line a density could be immediately calculated, using the equation presented in \cite{1996Rohlfs}:
\begin{equation}
\Sigma_\text{H{\sc i}} = 1.8\times10^{18} \text{cm}^{-2}/(\text{K km/s}).
\end{equation}
The mass of atomic gas as traced by H{\sc i} is found to be $5\times10^8\,\textrm{M}_\odot$, including a factor of 1.36 for He.

\subsection{Molecular Gas}

CO is traced using the CO(\textit{J}=2-1) rotational line data taken as part of IRAM's M33 Survey Large Program\footnote{\url{http://www.iram.fr/ILPA/LP006/}} \citep{2010Gratier,2014Druard}, which traces the molecular gas out to a radius of 7kpc using IRAM's Heterodyne arRAy (HERA, \citealt{2004Schuster}) instrument. This data has an angular resolution of 12\arcsec \,and a spectral resolution of 2.6km/s. The integrated intensity map can be seen in the right panel of Fig \ref{fig:gas_data}. We use this to trace molecular hydrogen within M33, rather that the earlier CO(\textit{J}=1-0) map \citep{2007Rosolowsky}, as this map only traces the CO out to a radius of 5.5kpc, and is less sensitive ($\sigma_\text{RMS}$=60\,mK for the (1-0) map, versus $\sigma_\text{RMS}$=20\,mK for the (2-1) data).

A conversion factor must be used to convert the CO intensity to a number density of $\text{H}_2$ ($X_{\text{CO}}$), and this value is uncertain. Historically, a value of approximately $2\times10^{20}\,\text{cm}^{-2}$ \citep{1987Solomon,1996StrongMattox,2012Smith} has commonly been used for the CO(\textit{J}=1-0) line, as calculated for the Milky Way, but a more comprehensive study by \cite{2013Sandstrom} found an average value of $1.42\times10^{20}\,\text{cm}^{-2}$ for a sample of 26 nearby galaxies. This conversion depends on a variety of factors, including the metallicity of the galaxy in question, and can vary across a galaxy. \cite{2012Narayanan} has shown that depending on the choice of $X_{\text{CO}}$, the Schmidt index can vary by $\pm 0.2$ and so the accurate treatment of this conversion factor is imperative. In this case, we used the values calculated by \cite{2010Braine} for M33. This work found two distinct populations: one within the central 2kpc of the galaxy ($X_{\text{CO}}=1.54\times10^{20}\,\text{cm}^{-2}$) and one outside this radius ($X_{\text{CO}}=2.87\times10^{20}\,\text{cm}^{-2}$). These values are for the CO(\textit{J}=1-0) line, so we turn these into conversion factors for the \textit{J}=2-1 line using the commonly employed ratio of $\text{CO} \left(\frac{2-1}{1-0}\right) = 0.7$ (e.g. \citealt{1990Eckart,2008Bigiel}). Including a factor of 1.36 for He, the total molecular gas mass was calculated to be $4.5\times10^7\,\textrm{M}_\odot$, an order of magnitude lower than the H{\sc i} mass.

\subsection{Gas traced by dust}

\begin{figure}
	\includegraphics[width=\columnwidth]{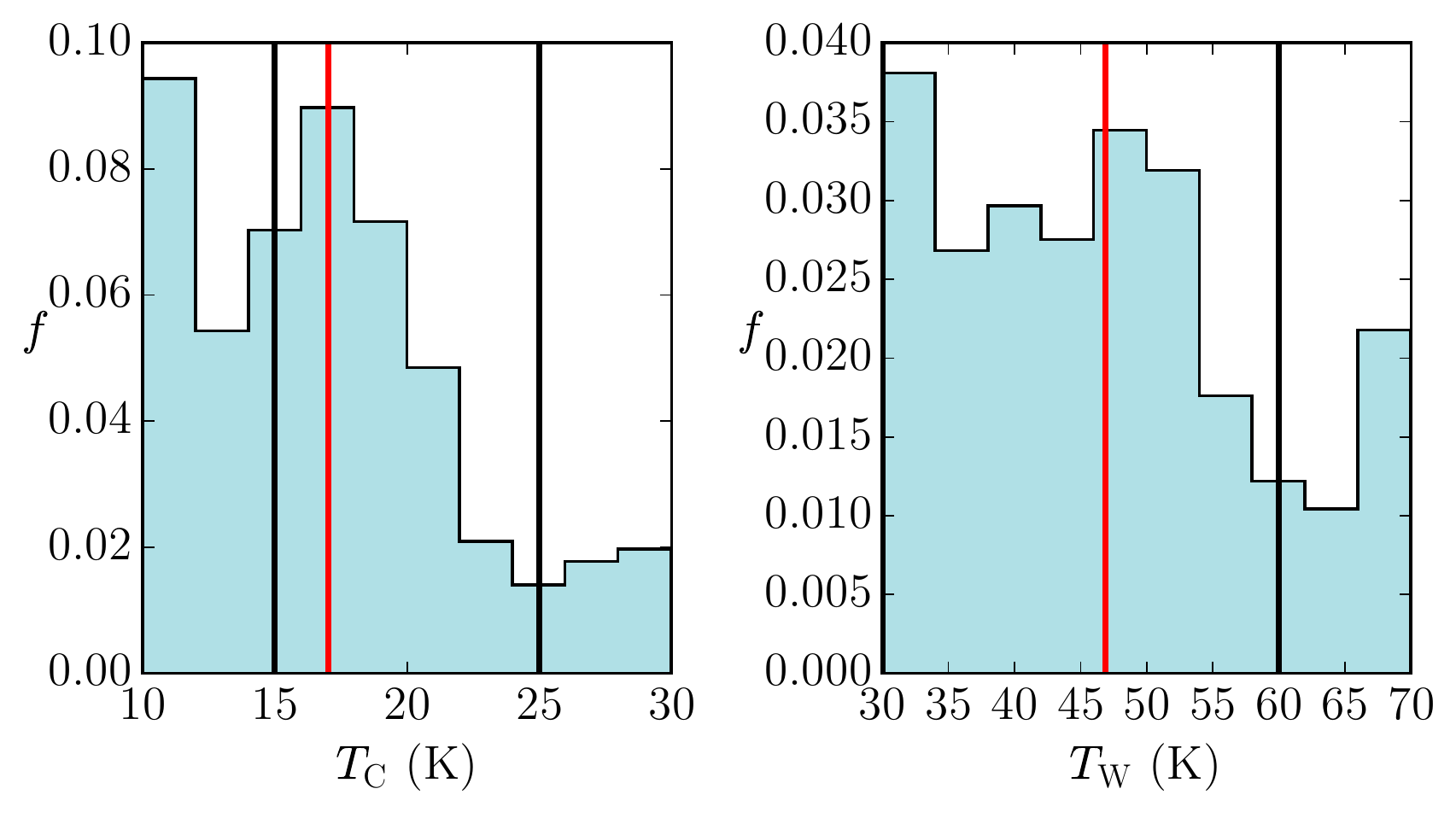}
    \caption{Distribution of cold and warm dust temperatures from MAGPHYS. The red line shows the average value of the sample, the black lines the bounds of the standard MAGPHYS parameter space.}
    \label{fig:t_distribution_magphys}
\end{figure}

\begin{figure}
	\includegraphics[width=\columnwidth]{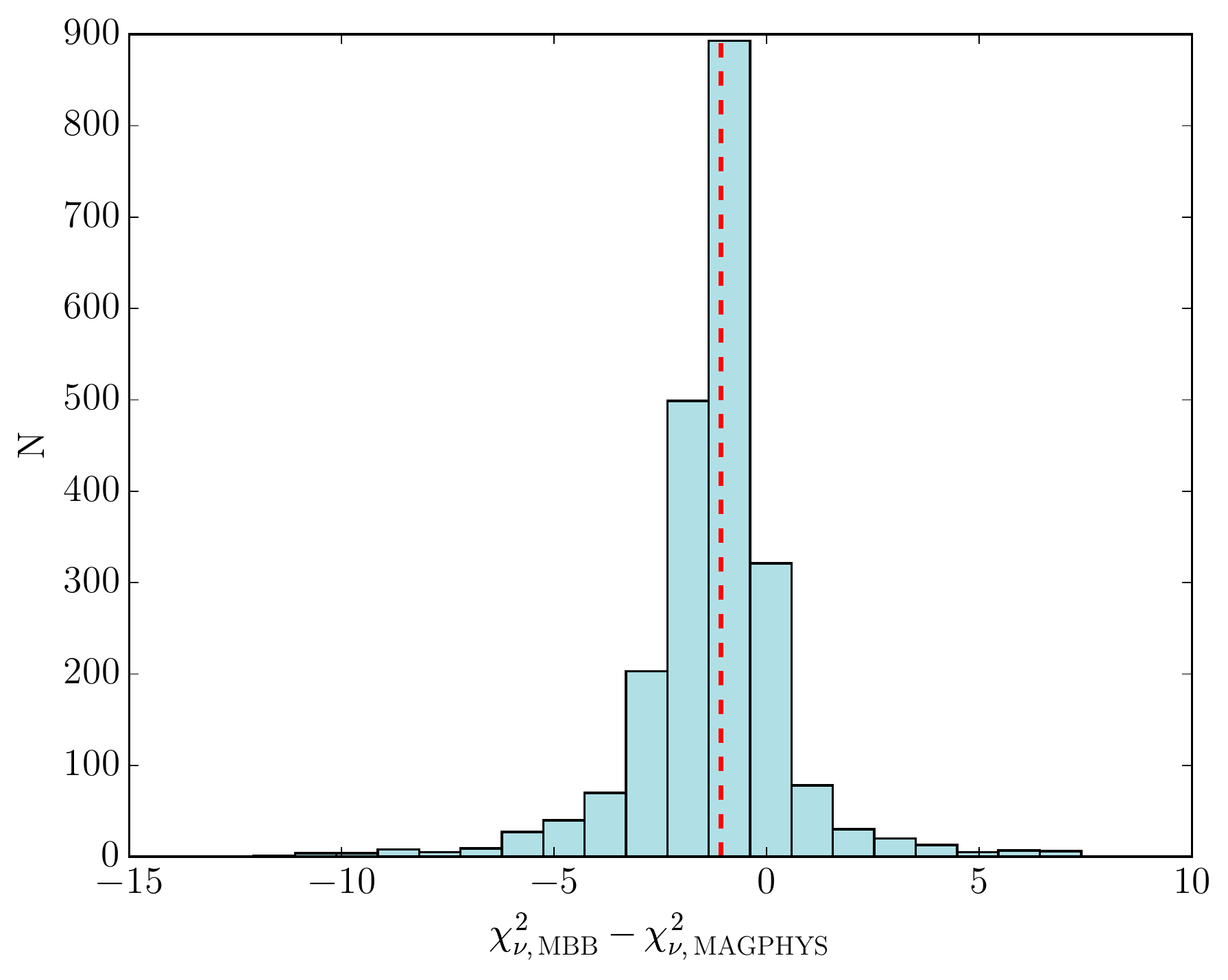}
    \caption{Comparison of reduced $\chi^2$ between MBB and MAGPHYS fits for individual pixels. The red dashed line indicates the median of the distribution (-1.08).}
    \label{fig:chi_sq_comparison_mbb_magphys}
\end{figure}

We also traced the gas via the dust in a galaxy, and to that end have created two dust maps. The first was a simple one-temperature modified blackbody (MBB), with a variable $\beta$, which takes the form
\begin{equation}\label{eq:mbb}
S_\nu = \frac{\kappa_\nu M_\text{dust}B(\nu,T_\text{dust})}{D^2},
\end{equation} 
where $S_\nu$ is the flux at frequency $\nu$, $\kappa_\nu$ is dust absorption coefficient at frequency $\nu$, $M_\text{dust}$ is the dust mass, $B(\nu,T_\text{dust})$ is the Planck function at frequency $\nu$ and dust temperature $T_\text{dust}$, and D is the distance to the source. We assumed $\kappa_{850} = 0.77\text{cm}^2\text{g}^{-1}$ \citep{2000Dunne}. Recent work by \cite{2016Clark} would suggest using a value of $\kappa_{500}=0.051^{+0.070}_{-0.026}\text{m}^2\text{kg}^{-1}$ instead, but we use this older value for consistency with MAGPHYS. We fit the MBB from 100-850\micron \,, using the 70\micron \,point as an upper limit to prevent fitting to warmer dust components \citep{2010Smith}. The dust mass and $\beta$ were allowed to vary freely, and the dust temperature was allowed to vary between 0-200K. This MBB fit was performed for all pixels with a S/N $> 2.5$ in at least 5 of the \textit{Herschel}/SCUBA-2 bands, giving us 2320 pixels with at least 1 degree of freedom, corresponding to a total area of 23.2kpc$^2$. The data has filter corrections suitable for extended sources applied, although no colour corrections were accounted for. Errors were accounted for via MCMC uncertainty estimation using \texttt{emcee}\footnote{\url{http://dan.iel.fm/emcee/current/}}. We used 100 walkers, each taking 300 steps, and the second half of these steps are used for error estimation to make sure the walkers have `burnt-in'. In each case, we take the error value to be the 16th and 84th percentiles of the samples in the marginalised distributions, and the 50th percentile as our value for the quantity in question. This MCMC estimation does not take into account correlated SPIRE uncertainties, but the effects of this are expected to be minor. This dust map can be seen in the left panel of Fig \ref{fig:dust_data}.

The second dust map was provided as an output by MAGPHYS. The MAGPHYS dust models incorporate polycyclic aromatic hydrocarbons (PAHs), which are based on observations of M17, and these features of the SED dominate at MIR wavelengths. MAGPHYS also includes hot dust, which it models with a series of grebodies with temperatures of 850, 250, and 130K. The warm dust is modelled with a modified blackbody with a emissivity index, $\beta$ of 1.5, and can vary between 30-60K. The cold dust is modelled in much the same way, but with $\beta=2$, and can vary between 15-25K. This map can be seen in the right panel of Fig \ref{fig:dust_data}.

Although these dust parameters are not unreasonable for an entire galaxy, at these sub-kpc scales some pixels may not fall within the standard MAGPHYS parameter space. We used an extended library of dust models \citep{2014Viaene} that increase the parameter space of the cold dust temperature from $10\text{K}<T_{\text{C}}<30\text{K}$, and the warm dust temperature to $30\text{K}<T_{\text{W}}<70\text{K}$. As with the SFR map, we performed filtering to remove any pixels for which the cold or warm dust temperatures were poorly constrained. The distributions of the cold and warm dust for these filtered pixels can be seen in Fig \ref{fig:t_distribution_magphys} -- with an average relative uncertainty on the cold dust temperatures of 4\% and on the warm dust of 10\%, $\sim$51\% of pixels are estimated to have cold dust temperatures outside of the standard MAGPHYS priors, with $\sim$32\% of pixels estimate to have warm dust temperatures outside of the standard range. It appears there may be a dust population with T < 10K, although this is most likely due to MAGPHYS using a fixed $\beta$ of 2 -- \cite{2012Xilouris} find that with a fixed $\beta$ of 1.5, the minimum dust temperature is 11K. This could also be due to the sub-mm excess in M33 \citep{2016Hermelo} causing MAGPHYS to fit colder dust components. However, performing the fits without the 850\micron \,data (where this submillimetre excess is most apparent) produced very little change in the cold dust temperatures. It also appears from the right panel of Fig \ref{fig:t_distribution_magphys} that a significant number of pixels are hitting the lower bound of the priors for warm dust temperature. This would indicate that these pixels are well fitted by a single-temperature blackbody fit, but as MAGPHYS enforces two temperatures to be fitted, this effect is unavoidable. A comparison of the reduced $\chi^2$ of the MBB and MAGPHYS fits (Fig \ref{fig:chi_sq_comparison_mbb_magphys}) finds a median offset of -1.08, indicating that on average, the MAGPHYS fits tend to be slightly worse.

\begin{figure}
	\includegraphics[width=\columnwidth]{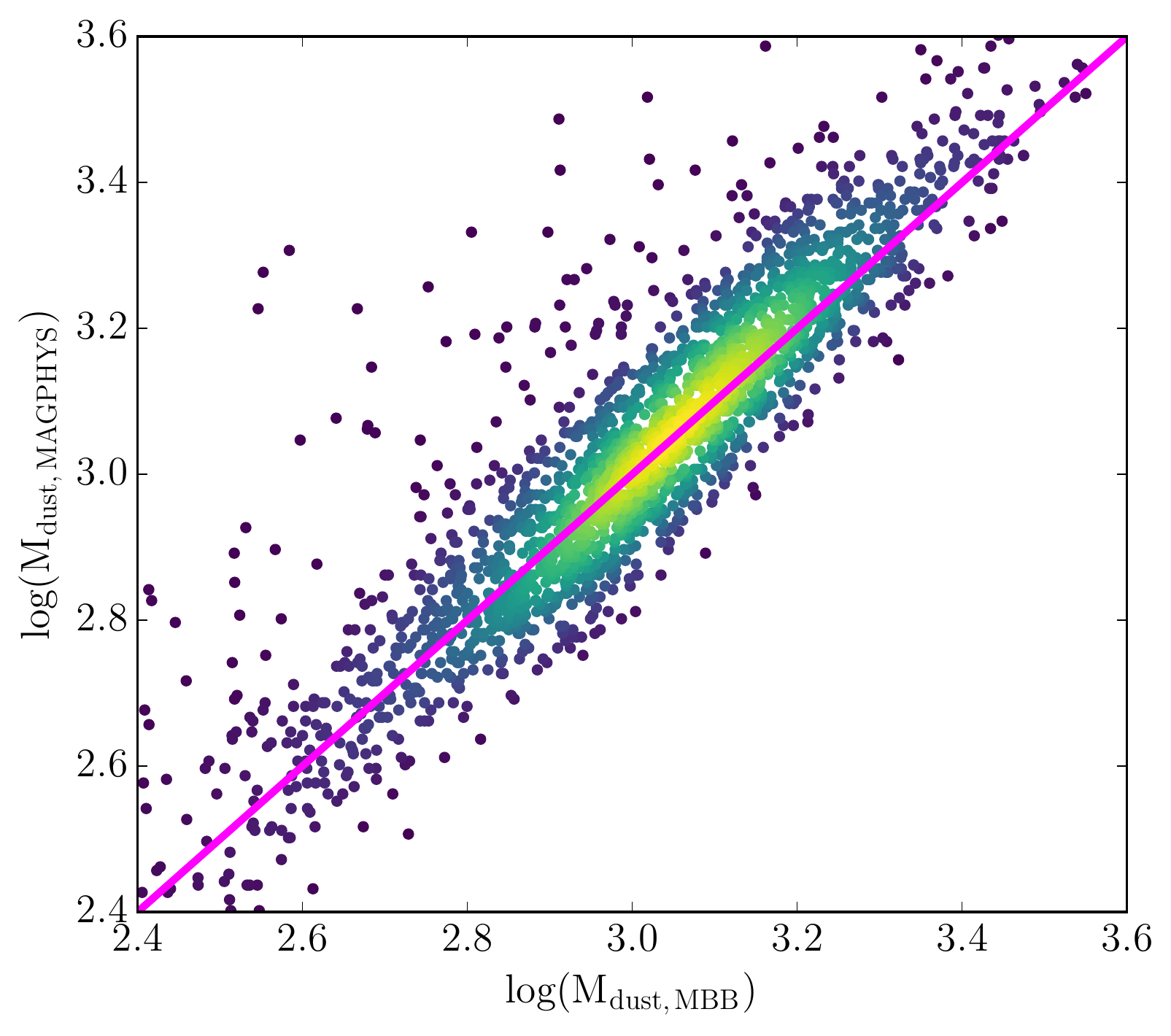}
    \caption{Comparison between the dust masses from MAGPHYS and MBB fitting. The one-to-one relation is shown as a fuchsia line. Blue points indicate a lower point density, yellow a higher point density.}
    \label{fig:dust_comparison}
\end{figure}

\begin{table*}
 \caption{Correlation coefficients between the various gas maps and the MAGPHYS SFR. All correlations have $p$ < 0.025.}
 \label{tab:gas_sfr_correlations}
 \begin{tabular}{cccccc}
  \hline
   & H{\sc i} & H$_2$ & Total gas & Total gas from single-temp blackbody & Total gas from MAGPHYS\\
  $\rho_\text{sp}$ & 0.35 & 0.57 & 0.42 & 0.44 & 0.51 \\
  $\rho_\text{pears}$ & 0.29 & 0.55 & 0.37 & 0.41 & 0.51 \\
  \hline				
 \end{tabular}
\end{table*}

A comparison of these two dust maps is shown in Fig \ref{fig:dust_comparison}. The two methods yield very similar results, and a very tight relationship with an RMS scatter of 0.10. There is a median offset of 0.02, indicating that the MAGPHYS calculated dust masses tend to be slightly higher. This is not unexpected, as the MBB has been fitted with a variable $\beta$. Many of the fitted $\beta$ values are less than 2 (the value MAGPHYS uses for its cold dust temperature), which will result in the MBB fitting higher dust temperatures, and thus lower dust masses. MAGPHYS takes into account a variety of dust compositions and temperatures, but this more sophisticated modelling does not greatly affect the calculated dust masses. As there is only a slight deviation from the one-to-one relation, this indicates that the MAGPHYS warm dust temperatures often running up against the lower bounds of the parameter space is not having a significant impact on the calculated dust masses.

The total gas mass can be calculated from dust masses using a dust to gas ratio (DGR). \cite{2013Sandstrom} find that the DGR is well correlated with metallicity, with a relationship given by
\begin{equation}
\log(\text{DGR}) = 0.55[12+\log(\text{O/H})]-6.50.
\end{equation}
We combined this with work from \cite{2008RosolowskySimon}, who find the metallicity gradient within M33 to take the form
\begin{equation}
12+\log(\text{O/H}) = -0.027R + 8.36
\end{equation}
where $R$ is in kpc. Combining these two, the radial variation in DGR is
\begin{equation}
\log(\text{DGR}) = -0.015R - 1.902.
\end{equation}
We note this gives similar results for the gas-to-dust ratio as the MW ($\sim$100, \citealt{1978Spitzer}), rather than the much higher results of 200-400 found in M33 by \cite{2017Gratier}. We find total gas masses of and 2.25$\times$10$^8$\,M$_\odot$ (MBB fits) and 9.75$\times$10$^8$\,M$_\odot$ (MAGPHYS).

\subsection{Which Gas Tracer best Correlates with SFR?}

To find out which of these gas maps best traced the SFR, we performed a series of Spearman's rank and Pearson correlation coefficient tests between these gas maps and the MAGPHYS calculated SFR. We also included a total gas map, combining the H{\sc i} and CO(\textit{J}=2-1) data. If the star-formation law breaks down at scales of $\sim$100pc, we would expect only a weak correlation here, so we regridded this data to 50\arcsec \,pixels (corresponding to 200pc), to mitigate against this effect. 

The correlations between the MAGPHYS SFR and the various gas maps can be seen in Table \ref{tab:gas_sfr_correlations}. We find that the correlations between gas and SFR are weaker than those found when comparing integrated galaxies (which typically have $\rho_\text{sp}\sim$0.8). This indicates much more scatter on these sub-kpc regions. Of the line-based gas tracers, we find that molecular gas has the strongest correlation -- this is not surprising, as work by, e.g., \cite{2008Bigiel} has also shown that molecular gas correlates better with SFR than the atomic gas, or the sum of the two.

Of the total gas traced by dust, we find that the MAGPHYS dust fits correlate better than the modified blackbody dust map. This is probably due to MAGPHYS more effectively tracing the total dust continuum, whereas the blackbody is only fitting to the cold dust. We elect to perform our analysis on the molecular gas, the total gas, and the total gas traced by dust from MAGPHYS fitting going forward.

\section{The Star-Formation Law}\label{sec:sf-law}

In this section, we investigate the star-formation law within M33 at high resolution. We start by comparing the average SFR and gas density to that of other galaxies in previous studies, before performing pixel-by-pixel fits within the galaxy. At high resolution, we also look at the radial variation in the Schmidt index. We have also investigated the effect of the spatial resolution used in the calculation of the KS index, as well as depletion timescale. Finally, we look at the relationship between SFR and dense gas at high resolution in M33.

\subsection{Global fits}

\begin{figure}
\begin{center}
\includegraphics[width=\columnwidth]{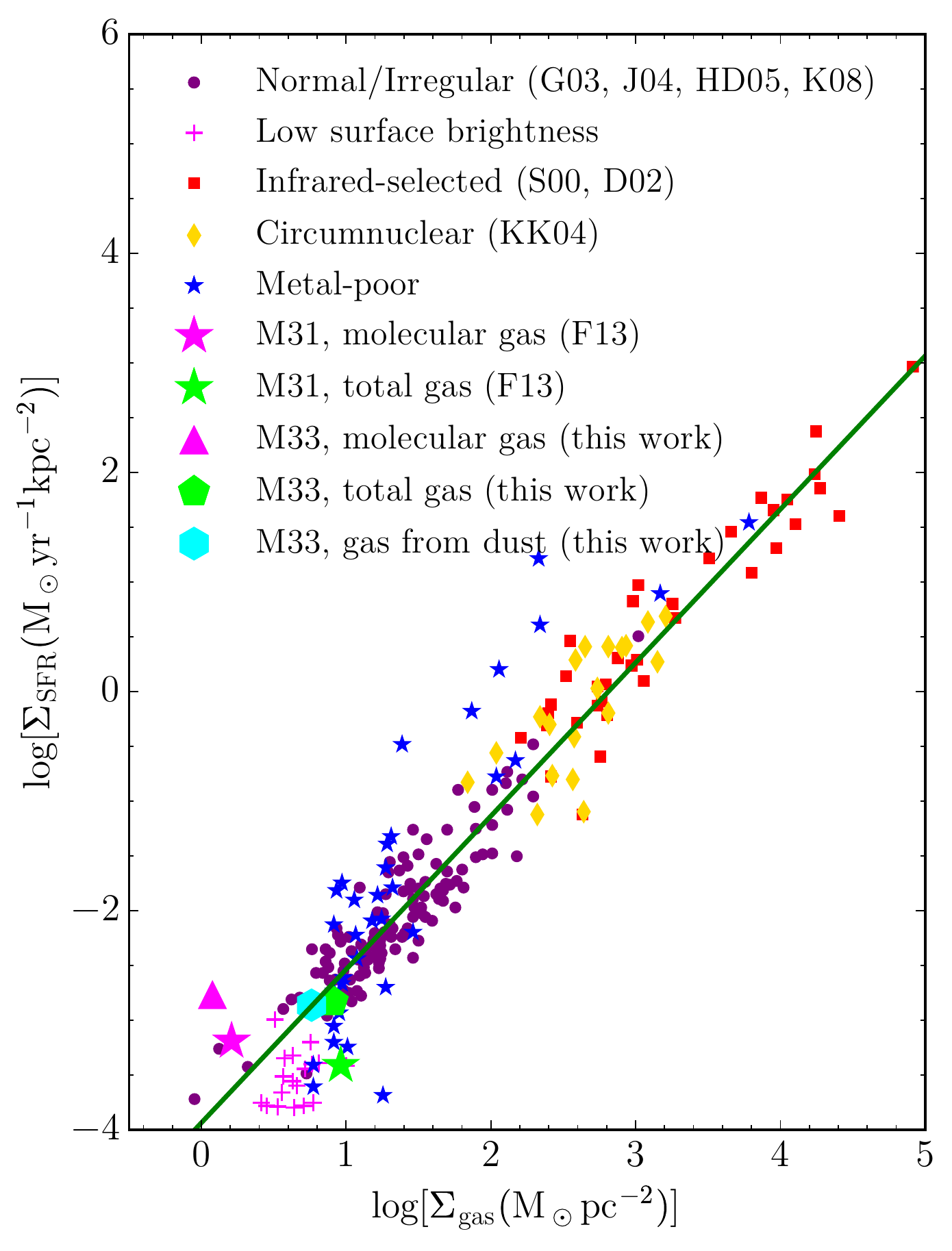}
\caption{Surface density of SFR against surface density of gas for a variety of galaxies. Purple circles indicate the normal and irregular galaxies of \citet{2003Gavazzi}, \citet{2004James}, \citet{2005HameedDevereux}, and \citet{2008Kennicutt}. Fuchsia pluses show the subset of these galaxies classified as low-surface brightness. The red squares are infrared-selected starburst galaxies of \citet{2000Scoville} and \citet{2002Dopita}, with circumnuclear starburst galaxies from \citet{2004KormendyKennicutt} shown as yellow diamonds. The subset of all of these galaxies with metal abundances, $Z<0.3Z_\odot$ are shown as blue stars. The values for molecular and total gas for M31 \citep{2013Ford} are shown as purple and light green stars, respectively. The purple triangle, light green pentagon and light blue hexagon show the values calculated in this work from the molecular gas, total gas, and gas from dust respectively. Also shown is a green line with slope, $N=1.4$.}
\label{fig:global_ks}
\end{center}
\end{figure}

Fig \ref{fig:global_ks} shows the mean surface density of SFR and gas for M33 compared with various other galaxies. The mean values here have been calculated from any pixels in the relevant gas map that match up with a pixel from the filtered MAGPHYS SFR map. In the case of the molecular gas, the surface density of SFR is about an order of magnitude higher than expected from this relationship. It is, however, consistent in terms of its molecular gas surface density with work by \cite{2013Ford} on M31, with this higher $\Sigma_\text{SFR}$ due to its higher star-formation rate and smaller size. The total gas, and gas from dust values lie on the trend. This is unsurprising, as literature values compare SFR surface density with total, rather than molecular gas. Globally, these values are similar to other galaxies, with a somewhat higher star-formation efficiency than M31. Overall, we find that the total surface density of gas and SFR is consistent with previous studies -- hence, M33 is a typical galaxy in terms of these parameters.

\subsection{Pixel-by-pixel fitting}\label{sec:pixel_by_pixel}

\begin{figure*}
\begin{center}
\includegraphics[width=1.8\columnwidth]{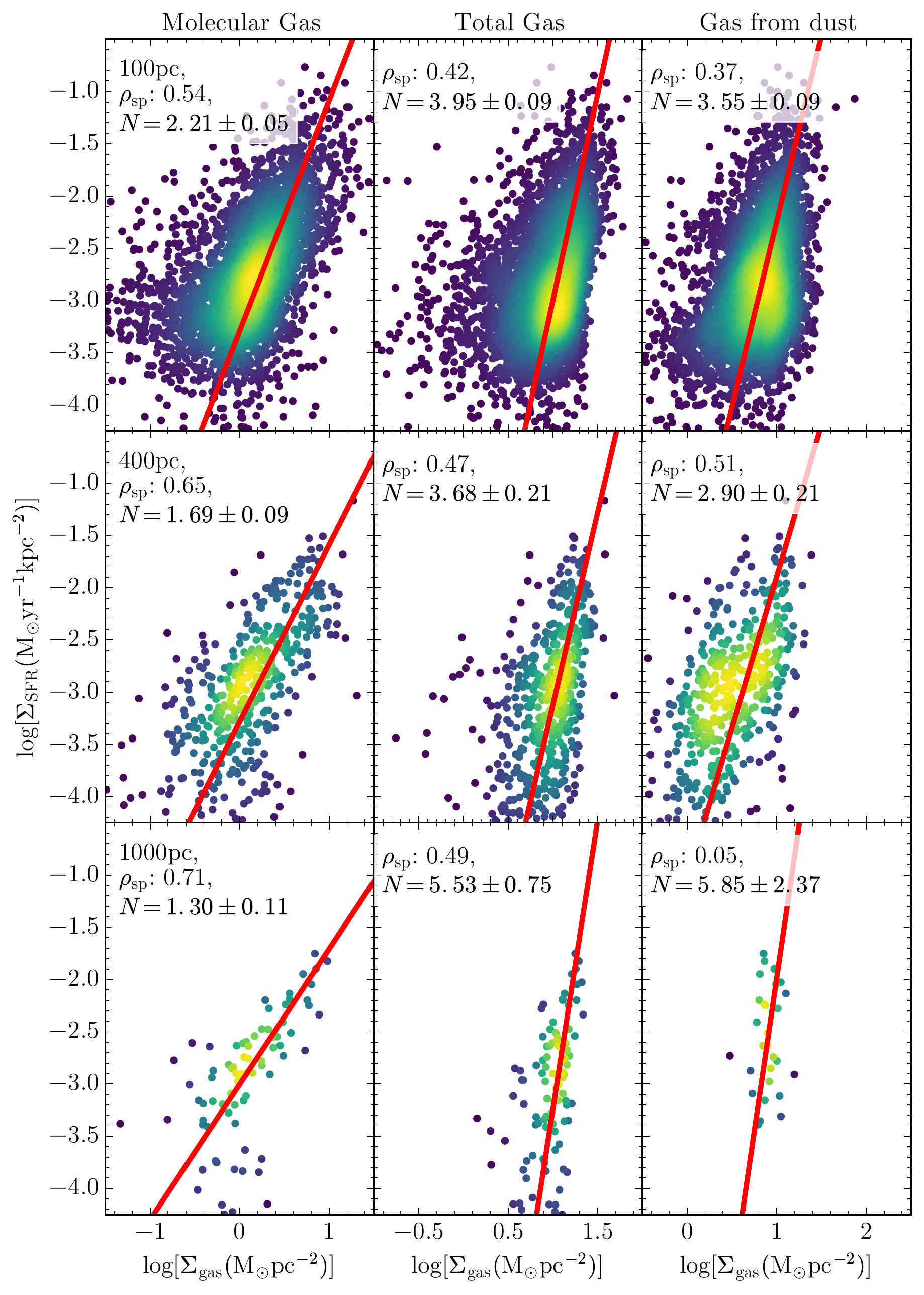}
\caption{\textit{Left column:} SFR surface density against molecular gas surface density. \textit{Middle column:} SFR surface density against total gas surface density. \textit{Right column:} SFR surface density against gas surface density, as traced by dust. The points are coloured by point density, from blue (low density) to yellow (high density). The red line indicates the best fit in each case.}
\label{fig:ks_fits_subkpc}
\end{center}
\end{figure*}

\begin{figure*}
\begin{center}
\includegraphics[width=1.9\columnwidth]{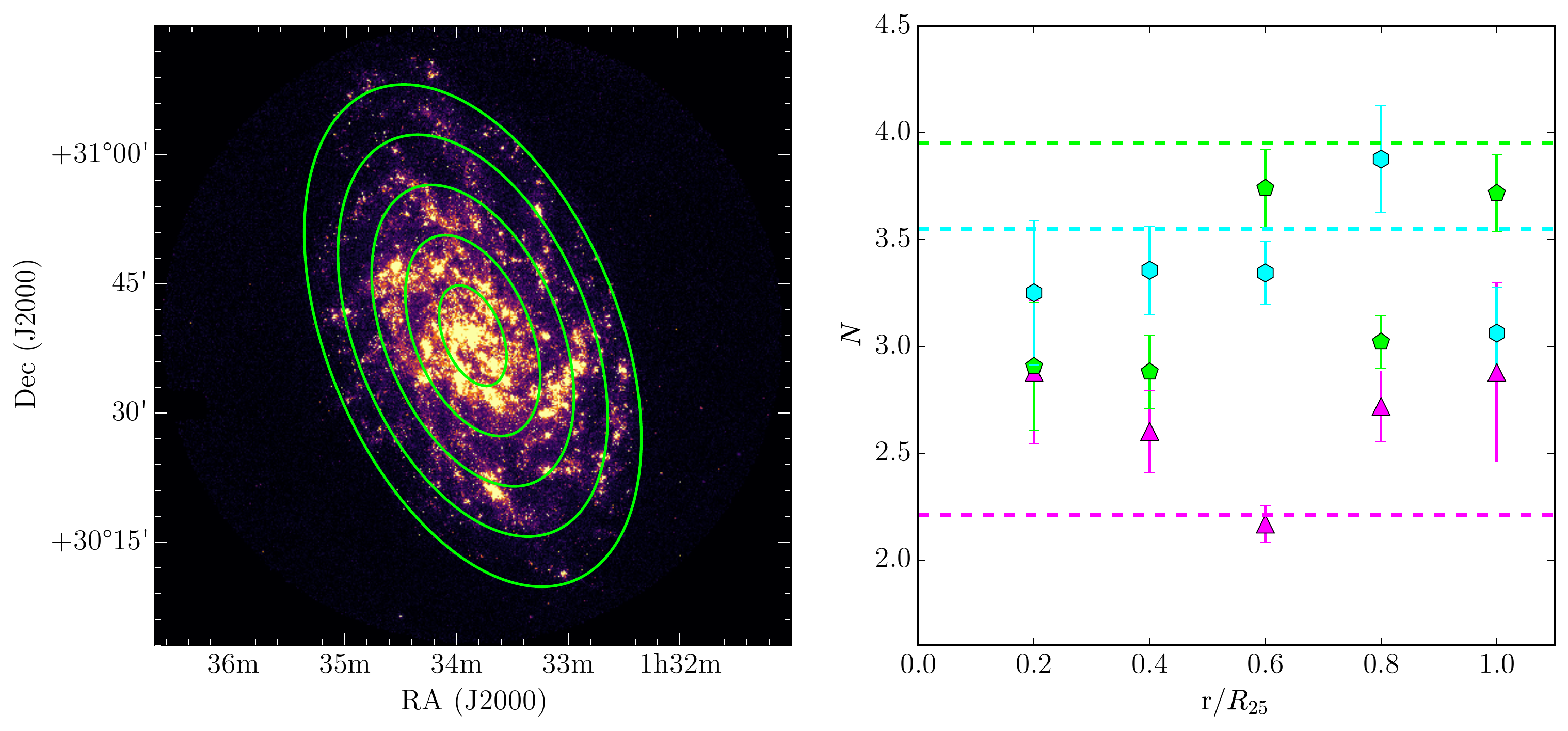}
\caption{Schmidt index with galactocentric radius within M33. Purple triangles indicate values calculated using the molecular gas, light green pentagons from the total gas, and light blue hexagons from the gas traced by dust. Also shown are dashed lines indicating the average value of $N$ fitted to all pixels.}
\label{fig:annuli_fits}
\end{center}
\end{figure*}

We also investigated the star-formation law on a pixel-by-pixel basis in M33. For this, we used all pixels in each gas map that matched up with a pixel in the SFR map, performing no S/N cut on the gas data, although by filtering the SFR map, we have effectively performed a S/N cut on the SFR map. This SFR cut biases the data somewhat, although we find that repeating the fits with the unfiltered data produces results consistent within the error bars. With pixel sizes of 100pc$\times$100pc, we converted the SFR and gas maps into units of surface density. These points can be seen for our three gas tracers in the top row of Fig \ref{fig:ks_fits_subkpc}. At spatial scales of 100pc, although we see significant scatter in the points, we find correlations between the gas and SFR surface densities ($\rho_\text{sp}$ = [0.54,0.42,0.37] for molecular gas, total gas and gas from dust respectively). This is in stark contrast to, e.g., \cite{2010Onodera,2010Schruba}, who find very little correlation between these quantities at comparable spatial scales. We attribute this to the methods these works have employed to calculate the SFR in these sub-kpc regions. These works calculated SFR using methods that assume constant star-formation over the last 100Myr, which the MAGPHYS fits and work by \cite{2009Relano} show is likely inappropriate over these sub-kpc regions. This can affect the calculated SFR by an order of magnitude (Fig. \ref{fig:sfr_comparisons}). We find that using our other SFR measures results in much more scatter at the highest resolutions, similar to that shown in \cite{2010Onodera}.

Another difference in our method is the choice of molecular gas tracer. Whilst \cite{2010Onodera} and \cite{2010Schruba} opt to use the {\it J}=1-0 line to trace the molecular gas, we use the {\it J}=2-1 line. Higher rotational transitions of CO trace warmer, denser, molecular gas, which may be more closely associated with star-formation. Work has shown that higher rotational transitions are more strongly correlated with SFR (e.g. \citealt{2007Komugi,2009Bayet}), and so this stronger correlation would be expected.

To calculate a Schmidt index, we use orthogonal distance regression (ODR), to account for errors in both $\Sigma_\text{SFR}$ and $\Sigma_\text{gas}$. The error in $\Sigma_\text{SFR}$ comes from the MAGPHYS modelling uncertainties, and the errors in gas surface density are derived from the RMS noise of the particular map (ignoring any calibration error, as this will simply cause an offset to all points and not affect the slope). In the case of the MAGPHYS dust map, we use the relative error calculated from the percentiles of the PDF. We perform this fit in linear space, as these error bars will be asymmetrical in log space. We fit this line to the entire data set, and these fits can be seen in Fig. \ref{fig:ks_fits_subkpc}. With the large range of S/N values of our $\Sigma_\text{SFR}$ and $\Sigma_\text{gas}$, particularly the large number of low S/N points, it is necessary to account for uncertainties in both variables when performing the model fitting. The effects of assuming a constant fractional uncertainty for all the data points rather than our measured uncertainty estimates is discussed in Appendix \ref{sec:app_const_err}.

At the highest resolutions, we find three very different indices to the classic $N=1.4$. Even in the case of molecular gas where we see the strongest correlations, the Schmidt index is higher than we would expect from \cite{1998Kennicutt}. To our knowledge, these values are higher than others previously reported. However, with strong correlations remaining between $\Sigma_\text{SFR}$ and the molecular gas, we argue that the star-formation law holds at these small scales. It also appears that the dust traces the total, rather than molecular gas.

\subsection{A search for a radial variation in $N$}

Work such as \cite{2008Leroy} and \cite{2013Ford} have shown that a radial variation can be seen in the Schmidt index in some galaxies. To investigate this at high resolution in M33, we have taken the pixels inside five annuli of constant galactocentric radius from the centre of M33 to $1.2R_{25}$. The results of this can be seen in Fig. \ref{fig:annuli_fits}. We see little radial variation in $N$ for all three tracers of gas, indicating that the star-formation efficiency is reasonably constant across the disk of M33. These results for all three gas tracers appear very different to those of \cite{2008Leroy} for spiral galaxies in general and \cite{2013Ford} for M31 in particular. We find that the calculated value of $N$ is reasonably consistent with each tracer of gas for these radial bins, with a peak in the outer spiral arms for the total gas, and the gas from dust.

\subsection{Variation with pixel scale}\label{sec:pix_scale_variation}

\begin{table*}
 \caption{Schmidt index for the molecular gas, total gas, and gas from dust, for a variety of pixel scales. Asterisks indicate correlations with $p$ > 0.025.}
 \label{tab:pixel_scale_global_fits}
 \begin{tabular}{lcccccc}
  \hline
  & \multicolumn{2}{c}{Molecular gas} & \multicolumn{2}{c}{Total gas} & \multicolumn{2}{c}{Gas from dust}\\
  Scale (pc) & $N$ & $\rho_\text{sp}$ & $N$ & $\rho_\text{sp}$ & $N$ & $\rho_\text{sp}$\\
  \hline
  100 & $2.21\pm0.05$ & 0.54 & $3.95\pm0.09$ & 0.42 & $3.55\pm0.09$ & 0.37 \\
  200 & $1.90\pm0.06$ & 0.57 & $1.93\pm0.12$ & 0.42 & $3.13\pm0.09$ & 0.51 \\
  400 & $1.69\pm0.09$ & 0.65 & $3.68\pm0.21$ & 0.47 & $2.90\pm0.21$ & 0.51 \\
  600 & $1.51\pm0.09$ & 0.70 & $3.90\pm0.29$ & 0.52 & $2.78\pm0.26$ & 0.57 \\
  1000 & $1.30\pm0.11$ & 0.71 & $5.53\pm0.75$ & 0.49 & $5.85\pm2.37$ & 0.05* \\
  2000 & $1.07\pm0.16$ & 0.87 & $5.20\pm1.33$ & 0.50 & $0.07\pm0.36$ & 0.14* \\
  \hline				
 \end{tabular}
\end{table*}

Although we find correlation at scales of 100pc, these are not as strong as those found by, e.g., \cite{2010Schruba} in M33 ($\rho_\text{sp}\sim$0.8 at 1200pc resolution). We expect this correlation to increase, and scatter between points to decrease with increasing pixel scale -- we average over GMCs in various evolutionary states at larger spatial scales, and it is this that is believed to drive the KS relation \citep{2010Onodera}. In order to test this increase in correlation, we regridded our data to a number of pixel scales (50\arcsec \,to 500\arcsec, 200pc to 2kpc). For the gas maps, we performed this using \texttt{Montage}'s mProject routine, and for any quantities derived from MAGPHYS we ran the fits on the regridded data. We then performed a fit to each pixel scale, using the same method as detailed in Section \ref{sec:pixel_by_pixel}. A selection of these fits can be seen in the lower panels of Fig \ref{fig:ks_fits_subkpc}, and the calculated values of $N$, along with Spearman's rank correlations can be seen in Table \ref{tab:pixel_scale_global_fits}.

Between pixel scales and gas tracers, we see variation in the Schmidt index. With molecular gas, this decreases with increasing pixel scale before becoming approximately linear at a scale of $\sim$2kpc. This value is consistent with the index found by \cite{2008Bigiel} when considering molecular gas, and indicates that at kpc resolutions, these GMC populations appear much more uniform, and we are simply counting the numbers of them. We also find that, in general, $N$ increases with increasing pixel scale for the total gas. There is, however, significant scatter in our calculated $N$ with pixel scale with gas from dust, and at large pixel scales the correlation between this tracer of gas and SFR is no longer statistically significant. For our more conventional gas tracers, we find an increasing correlation between the surface density of gas and SFR with increasing pixel scale, with decreasing scatter from the relationship. This would indicate that the star-formation law we see at integrated galaxy scales is driven by an average of GMCs at various evolutionary states in a galaxy.

\begin{figure}
\begin{center}
\includegraphics[width=\columnwidth]{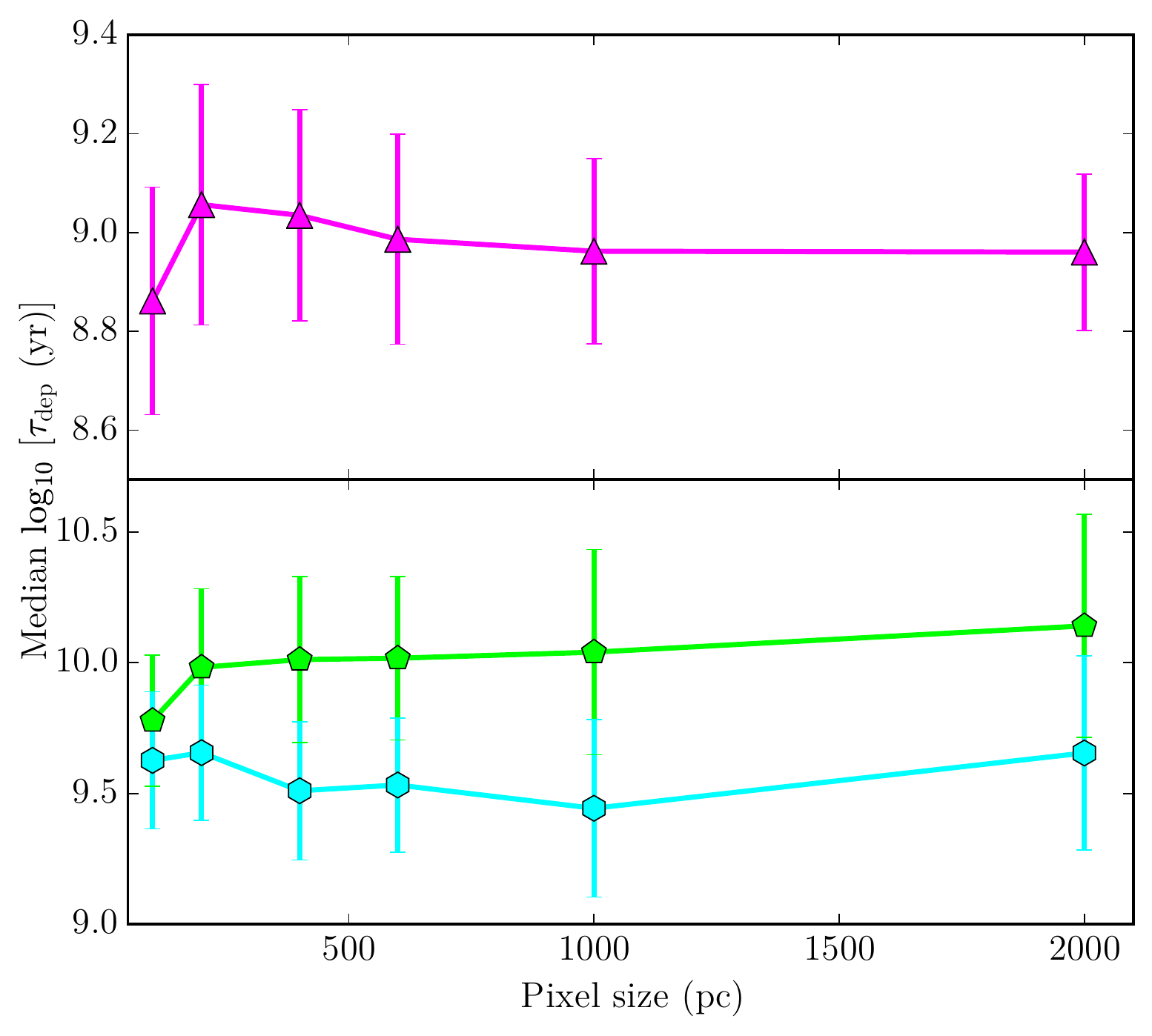}
\caption{Depletion timescales for the three gas tracers with pixel size. Purple triangles indicate the molecular gas, green pentagons the total gas, and light blue hexagons the gas traced by dust.}
\label{fig:timescale}
\end{center}
\end{figure}

We also investigate the scale dependence on the gas depletion timescale -- this can be seen in Fig. \ref{fig:timescale}. We calculate the depletion timescale in the same manner as \cite{2010Schruba}, where $\tau_\mathrm{dep} = \Sigma_\mathrm{gas}/\Sigma_\mathrm{SFR}$. We quote the $1\sigma$ errors based on the percentiles of the depletion timescale distribution. Unlike \cite{2010Schruba}, we find no significant variation in the gas depletion timescale with resolution, for any of our three gas tracers. This would appear to be due to their targeted selection of only the brightest regions of CO and H$\alpha$ -- when taking into account the entire ensemble of regions within M33, these resolution effects are no longer significant.

It is also important to note that these results cover a much smaller dynamic range than Fig. \ref{fig:global_ks}, and so the correlations would naturally be weaker. We took a subset of the data in Fig. \ref{fig:global_ks} over the $\sim$2 orders of magnitude that our data covers. We found that the average Spearman correlation coefficient is $\sim$0.84 for these data, still stronger than we find for M33 at any spatial scale. To recover the correlation we see comparing galaxy to galaxy, it seems necessary to take into account the entire ensemble of GMCs within that particular galaxy.

We find that these data also do not reproduce the correlations seen by \cite{2010Schruba}, who find a correlation coefficient of $\sim$0.8 at scales of $\sim$1kpc. However, their work used targeted apertures on CO and H$\alpha$ peaks, biasing their results towards areas of high star-formation and S/N, where this star-formation law holds more strongly. When we place apertures centered on peaks of gas or SFR, rather than pixel-by-pixel comparisons, we find a stronger correlation ($\rho_\text{sp}=0.82$ for apertures of 1200pc diameter), comparable with that of \cite{2010Schruba}. Our work blindly includes all areas within a galaxy, so we avoid the very high S/N requirements of \cite{2010Schruba}. Thus, this is the cause of these slightly weaker correlations.

\subsection{SFR and Dense Gas}

\begin{figure}
\begin{center}
\includegraphics[width=\columnwidth]{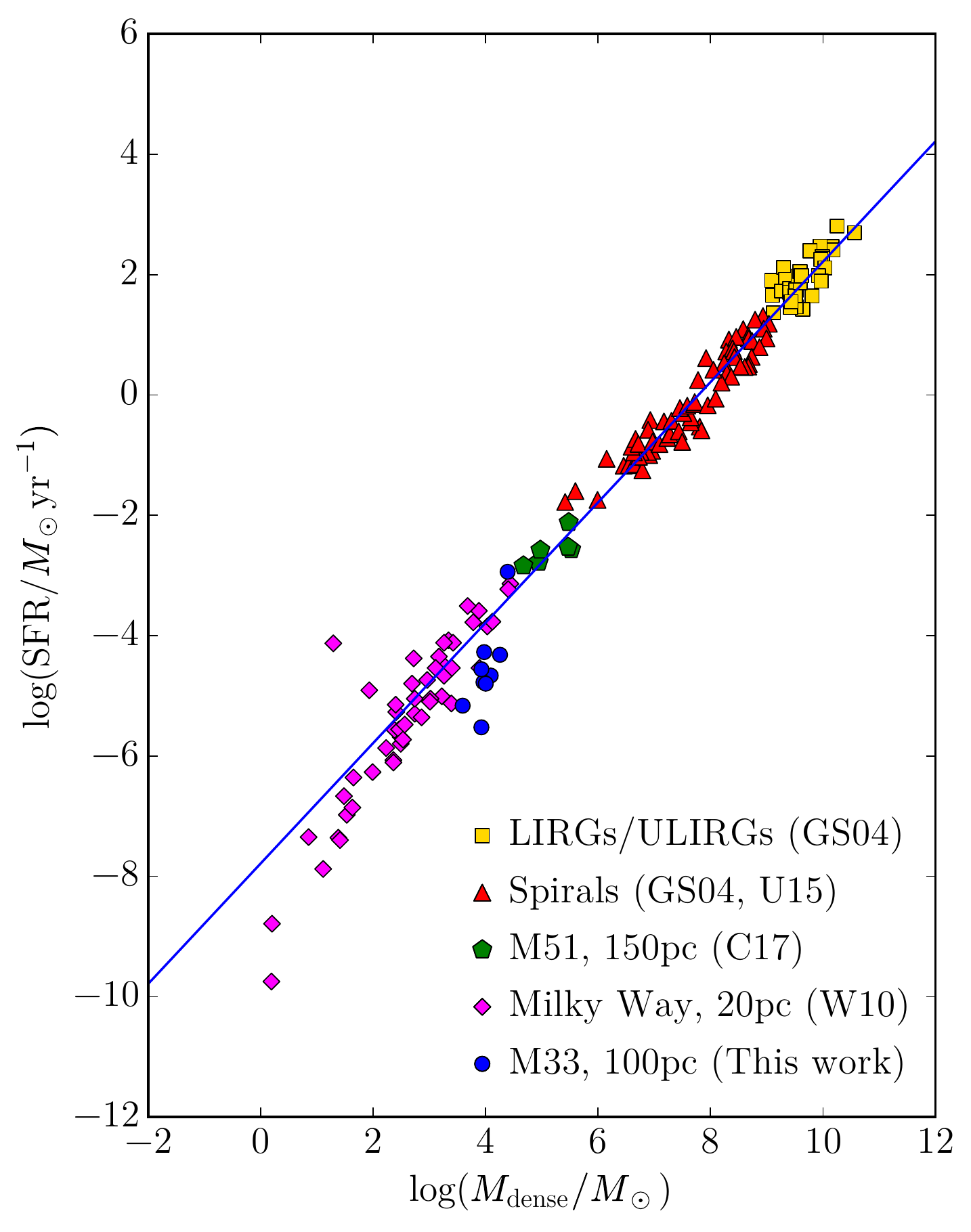}
\caption{Relationship between SFR and dense gas mass across a range of environments and scales. Yellow squares indicate values for the LIRGs and ULIRGs of \citet{2004GaoSolomonb}, red triangles spiral galaxies from the surveys of \citet{2004GaoSolomonb} and \citet{2015Usero}. Green pentagons are sub-kpc measurements taken in M51 by \citet{2017Chen}, purple diamonds are from the MW pointings of \citet{2010Wu}. Blue circles indicate values for this work -- SFR values have been calculated using MAGPHYS, and dense gas pointings come from a variety of sources, described in the main text. In every case, the dense gas mass has been calculated using HCN(\textit{J}=1-0), and for pointings that are resolved, the approximate spatial scale is given in the legend. The blue line indicates a linear fit to the data.}
\label{fig:dense_gas}
\end{center}
\end{figure}

We have investigated the relationship between dense gas and SFR, to see if this relationships show a similar scale dependence to the star-formation law. We have used HCN(\textit{J}=1-0) emission to trace the dense molecular gas. Most of these dense gas pointings come from \cite{2013Buchbender}, with some complementary measurements from \cite{2011Rosolowsky,2017Braine}. All of these pointings were from the IRAM 30m telescope, with a beam size of 28\arcsec, comparable to our pixel size. To convert the HCN luminosity to a mass of dense H$_2$, we used a conversion factor, $\alpha_\text{HCN}=10$ \citep{2004GaoSolomona}. The results of this can be seen in Fig \ref{fig:dense_gas} -- a strong correlation is seen between the dense gas mass and SFR ($\rho_\text{sp}=0.68$, $p=0.04$). Additionally, these points lie approximately on the extrapolation of the linear fit seen for integrated galaxies, indicating that this dense gas relation holds down to these 100pc scales. Possibly the reason these points are somewhat below the line is due to the low metallicity of M33, since due to their sensitivity to photodissociation, dense gas tracers are very strongly dependent on metallicity \citep{2011Rosolowsky}. Nevertheless, these results indicate that the dense molecular gas correlates more strongly than the regular molecular gas with SFR at these sub-kpc scales.

\section{Discussion and Conclusions}\label{sec:discussion_conclusions}

In this work, we have presented a high-resolution study of the star-formation law in M33 on spatial scales of $\sim 100$pc. By assembling GALEX UV, SDSS, WISE, \textit{Spitzer}, \textit{Herschel}, Planck, and new SCUBA-2 observations we have measured the SFR using the panchromatic SED fitting tool MAGPHYS. We have compared this SFR with that calculated both by the TIR luminosity and FUV+24\micron \,data. We find that much of the starlight in M33 is unattenuated by dust, leading to an underestimate of the total SFR from the TIR luminosity. MAGPHYS models many of the H{\sc ii} regions of M33 as starburst-like, and thus produces pixels with much higher SFR than even the FUV+24\micron \,calculated SFR. Since MAGPHYS uses all available data and allows variations in SFR to much shorter timescales than the TIR and FUV+24\micron \,prescriptions, we have used this in our analysis.

We have combined the SFR calculated from these SED fits with gas maps created from H{\sc i} and CO(\textit{J}=2-1) data. We have also constructed dust mass maps of M33 using both a MBB fitting code and MAGPHYS, and compared the two. We find that these two maps agree very closely, with the MAGPHYS masses tending to be slightly higher -- not unexpected with a variable $\beta$ and a single temperature in the case of the MBB fitting. It is important to note that when fitting a two-temperature MBB to M33, \cite{2014Tabatabaei} find that a fixed $\beta$ of 1.5 for the cold dust component was a better fit than the 2 the MAGPHYS uses, but the effect of this is minor. Using the DGR calculated from the metallicity gradient, we turn these dust maps into total gas mass maps. We use maps of the molecular gas, total gas (CO + H{\sc i}) and total gas from dust to probe the star-formation law at scales of 100pc.

We find that M33 is not an unusual galaxy in terms of its overall gas and SFR surface density, and whilst correlations remain down to scales of 100pc, the measured Schmidt index shows a strong scale dependence. This indicates that the GMCs within M33 are at a variety of evolutionary states, and so the star-formation law is very different at GMC, rather than galaxy scales. We also find that at these scales, molecular gas better traces SFR. The gas depletion timescale, however, shows no such scale dependence. We find that $N$ is reasonably invariant with galactocentric radius, with a peak in $N$ for total gas and gas from dust in the outer spiral arms. If we consider the dense gas mass of a galaxy, a tight, linear relationship is found, perhaps indicating that dense molecular gas is the fundamental building block of star-formation.

Using a wide range of high-resolution data, and leveraging the close proximity of M33, we have been able to probe the light at GMC scales in this galaxy across some four orders of magnitude in wavelength. From this broad range of coverage, a large range of galaxy parameters can be calculated, and various laws probed down to the small-scale. It would appear that, at the scales of GMCs, the star-formation law does hold, although the Schmidt index is very different at these scales. We also find a quasi-universal star-formation law with dense molecular gas.

\section*{Acknowledgements}

The authors would like to thank the reviewer for helpful comments, Stephen Eales and Jenifer Millard for valuable discussions, S{\'e}bastien Viaene for providing the extended MAGPHYS IR libraries, and Elisabete da Cunha for MAGPHYS assistance. T.G.W. would also like to personally thank Ken Fields. 

M.W.L.S. acknowledges support from the European  Research  Council  (ERC)  Forward  Progress  7 (FP7)  project  HELP. The James Clerk Maxwell Telescope is operated by the East Asian Observatory on behalf of The National Astronomical Observatory of Japan, Academia Sinica Institute of Astronomy and Astrophysics, the Korea Astronomy and Space Science Institute, the National Astronomical Observatories of China and the Chinese Academy of Sciences (Grant No. XDB09000000), with additional funding support from the Science and Technology Facilities Council of the United Kingdom and participating universities in the United Kingdom and Canada.

This research has made use of the NASA/IPAC Infrared Science Archive, which is operated by the Jet Propulsion Laboratory, California Institute of Technology, under contract with the National Aeronautics and Space Administration, as well as the NASA/IPAC Extragalactic Database (NED) which is operated by the Jet Propulsion Laboratory, California Institute of Technology, under contract with the National Aeronautics and Space Administration. 

This research made use of Montage, which is funded by the National Science Foundation under Grant Number ACI-1440620, and was previously funded by the National Aeronautics and Space Administration's Earth Science Technology Office, Computation Technologies Project, under Cooperative Agreement Number NCC5-626 between NASA and the California Institute of Technology. 

This research made use of Astropy, a community-developed core Python package for Astronomy (Astropy Collaboration, 2013).



\bibliographystyle{mnras}
\bibliography{bibliography}



\appendix

\section{Fitting With Uniform Weights}\label{sec:app_const_err}

\begin{figure}
\begin{center}
\includegraphics[width=\columnwidth]{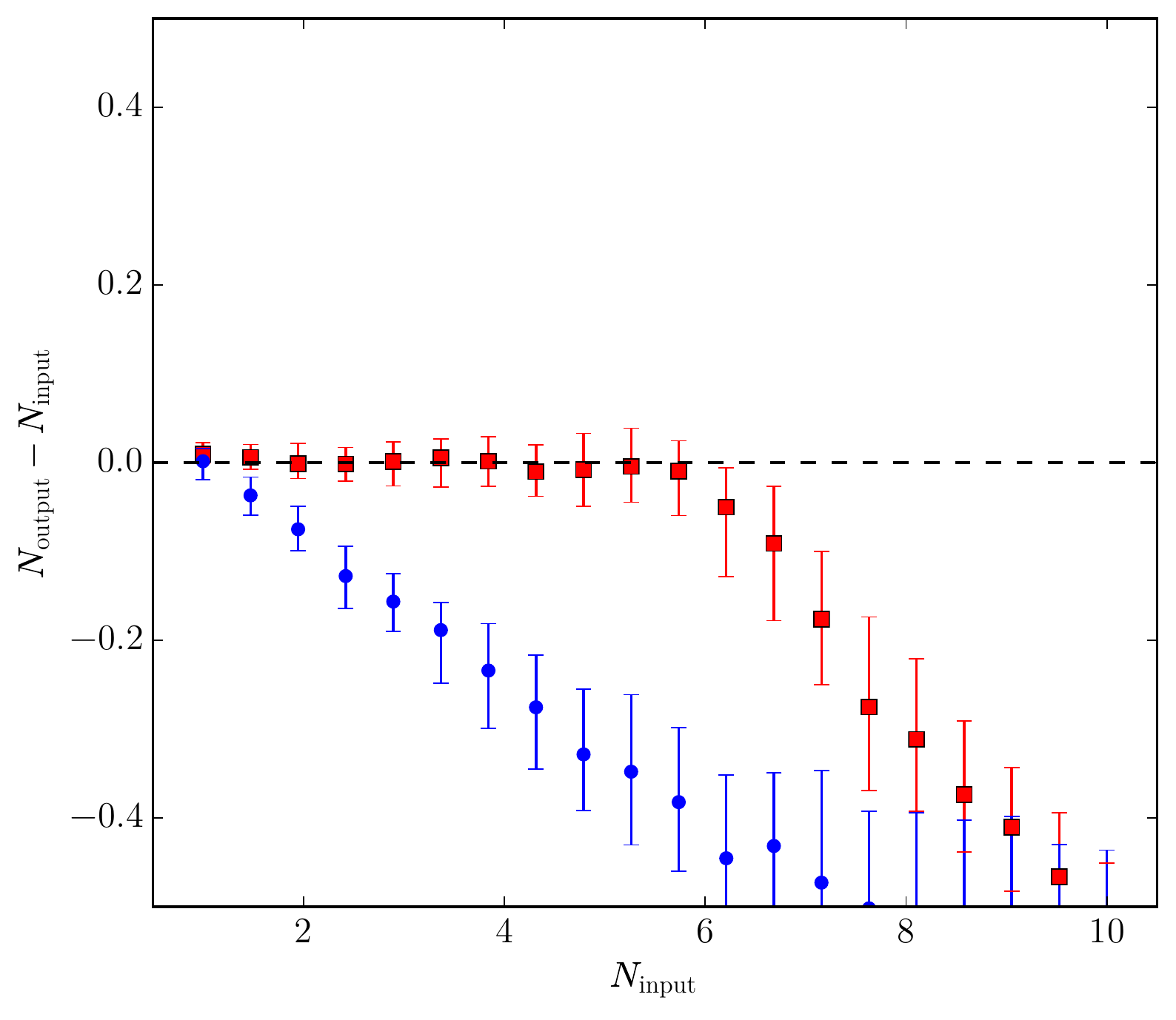}
\caption{Recovered $N$ against input $N$ for a variety of inputted slopes. The blue dots indicate points where we have assumed a constant percentage error in SFR and gas surface density, the red squares show the errors weighted based on the errors in the relevant maps. The dashed line shows where the recovered slope is equal to the input slope.}
\label{fig:fit_sim}
\end{center}
\end{figure}

To avoid biasing our data set unnecessarily, we fit to all pixels regardless of S/N. As many of these points (particularly with the molecular gas) are low S/N, it is important to take into account the errors on these points, and we highlight this here. We have performed a number of simulations where we sample 1000 points distributed normally over the dynamical range of the total gas density, and input a slope $N$ that ranges from 1 to 10. The points are given scatter in both $\Sigma_\text{SFR}$ and $\Sigma_\text{gas}$ based on the approximate scatter we see in the data (here, we have assumed normal distributions in both $\Sigma_\text{SFR}$ and $\Sigma_\text{gas}$). We performed a fit in two different ways -- firstly by weighting the errors assuming a constant percentage error on the data, and secondly by weighting by the RMS errors in our map. For each inputted $N$, we repeat this fit 100 times, and calculate the median recovered $N$, with the 16th and 84th percentiles forming our lower and upper errors. The results of these simulations can be seen in Fig \ref{fig:fit_sim}.

For any simulated $N$ above a value of 1, we see that the uniformly weighted error systematically underestimates the slope. When weighting based on the RMS errors of the map, we can reliably recover slopes up to an $N$ of 6. This is due to the percentage errors being larger at lower S/N, and this variation at these low values dominating the fit, especially as the gradient of the slope increases. Since the distribution of $\Sigma_\text{gas}$ in the real data is skewed to lower S/N rather than normally distributed, we would expect this effect to be even more pronounced in the real data. Realistically accounting for the errors in $\Sigma_\text{SFR}$ and $\Sigma_\text{gas}$ allows us to recover the underlying Schmidt index of the data across the entire range of $N$ that we calculate in our work, we opt to use this method, rather than assuming a uniform weighting in either $\Sigma_\text{SFR}$ or $\Sigma_\text{gas}$.


\bsp	
\label{lastpage}
\end{document}